# Implicit and Explicit Solvent Models for the Solubility of 1,4-Diethynylbenzene and Terephthalonitrile in Ionic Liquids

Clare Xie Yijia,† Julian Becker,† Tom Welton,† Patricia Hunt*
*†: Department of Chemistry, Imperial College London*
**:School of Chemical and Physical Sciences, Victoria University of Wellington*

## Abstract

The interactions of ionic liquids (ILs) with organic solutes are of interest to many applications involving the use of ILs as task-specific solvents, such as in sensing devices or catalysis. However, currently little is known about these specific interactions. The interactions of two structurally-similar organic analytes with different functional groups (alkyne and cyanide) in the IL $[C_2C_1Im][MeSO_4]$ are studied via density functional theory. By analysing the type and strength of the interactions between the organic solutes and ionic liquid using explicit solvation, the study aims to better understand of the possible noncovalent interactions between them. Implicit solvation models were also adopted to obtain solvation energies from solvent continuum. The understanding of IL-solute interactions and solubilities are critical in many applications such as improving the analytical performance of ILs-based sensors.

## Introduction

### *Introduction to ionic liquids and applications*

ILs are salts with melting points near room temperature.[1] ILs possess a unique combination of properties including low melting points, high thermal stability and high electrical conductivity.[2] This has led to a growing interest in IL applications in sustainable chemistry,[3] such as acting as green solvents in different industrial applications,[4] including catalysis, photovoltaic devices,[5] batteries,[1] and electrochemical sensors.[6] Common ILs are often formed by combining a larger organic cation and a simpler inorganic anion (**Figure 1**).[7]

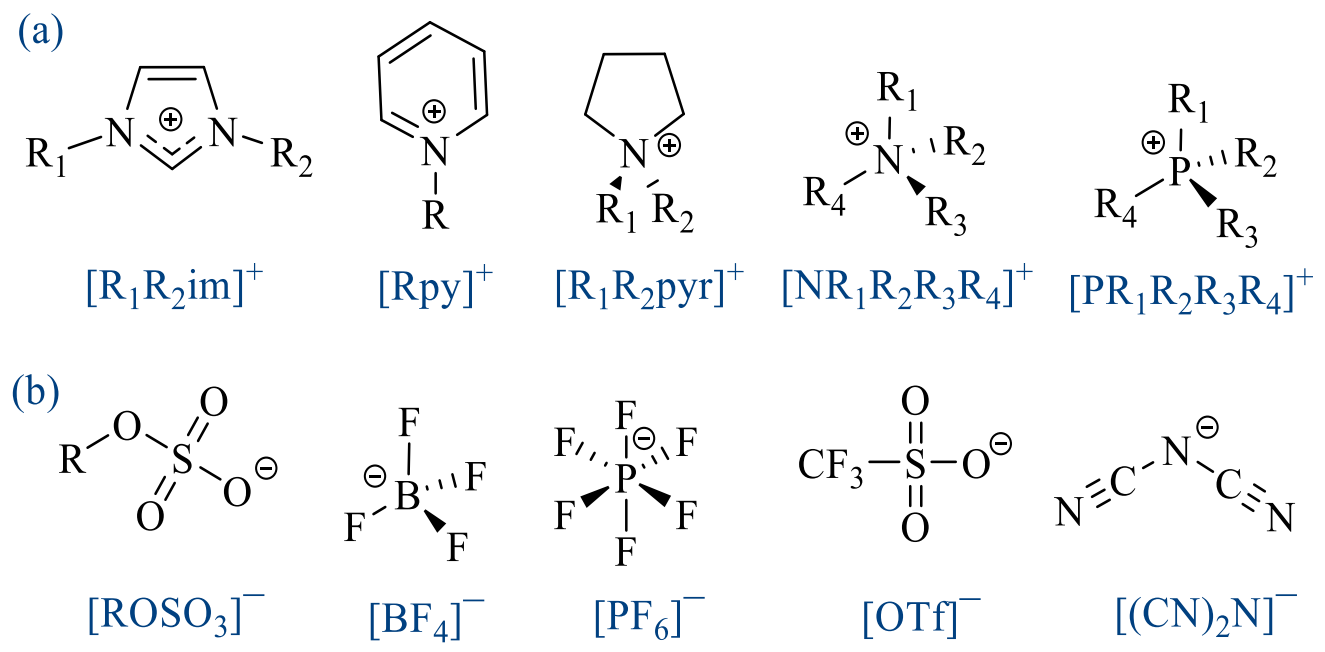

Figure 1: Cations (a) and anions (b) of common ILs.
(im: imidazolium, py: pyridinium, pyr: pyrrolidinium.)

Properties of ILs for specific applications can be fine-tuned through identifying suitable cations and anions from structural libraries and varying substitution patterns.[8] [7] However, due to the large number of possible combinations of cation and anions, it is impossible to investigate all potential ILs experimentally.[7] Computational methods offer an efficient way to obtain preliminary understanding of the intermolecular interactions of ILs and plays a crucial role in the identification of ILs for applications.

Electrochemical sensors rely on the solubility of the analyte in the electrolyte to detect the analyte's presence.[9] Solubility is a thermodynamic term used to describe the ability of the solute to dissolve in the solvent. The dissolution process involves solvent and solute reorganization to maximize attractive solvent-solute interactions.[10] Saturation is a measure of solubility and is achieved when adding more solute to the solution will not increase the concentration of the solute further, and the solution system is at thermodynamic equilibrium. Therefore, ILs used as solvents should have the correct electronic and solvent properties to form favourable interactions with the intended solute.[1] [11] The Gibbs free enthalpy of solution ($\Delta G_{solution}$) describes how favourable the dissolution process is.[12] In this study, the solvation process from the gaseous state to solution was examined computationally, and the solvation energies ($\Delta G_{solvation}$) were calculated to understand $\Delta G_{solution}$.

Imidazolium-based ILs contain an imidazolium cation with the single positive charge on the $C^2$ carbon and on the two nitrogen atoms of the ring (**Figure 2**).[13] Imidazolium cations can be abbreviated as $[C_nC_mIm]^+$, where n and m are the number of carbons in the alkyl chains.[14] These ILs can be used as reaction media for organic synthesis,[15] catalysis,[16] and pre-treatment of coal for better gasification,[17] or even as absorbents in carbon capture.[18] Because of the wide electrochemical windows, imidazolium-based ILs are a good choice for electrolytes in electrochemical sensors and supercapacitors.[19]

ILs containing an alkyl sulfate anion (**Figure 2**) also have desirable electrochemical and thermal stabilities [20] which makes these ILs promising electrolytes for use in electrochemical devices[21] and catalysis.[22] Ethylsulfate-containing ionic liquids were found to exhibit wide electrochemical windows of about 5.0 V.[21] The cations in ionic liquids containing alkyl sulfate anion have shown to greatly influence viscosity, density, and conductivity.[22] The particular imidazolium-based alkyl sulfate ILs examined in this study were found to be favourable in applications for desulfurization,[23] catalysis,[24] and gas sensors.[25]

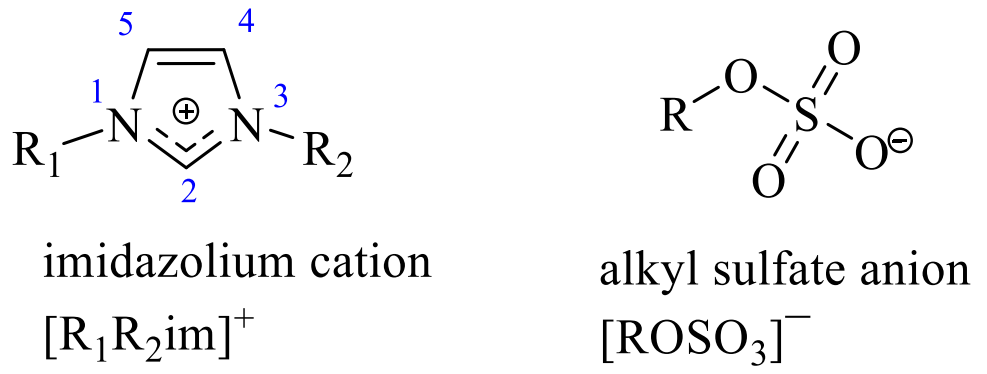

Figure 2: General structures of the imidazolium cation and alkyl sulfate anino.

The interactions within ILs are mainly noncovalent interactions (interactions that do not consist of strong directional covalent bonds). These interactions include ionic, hydrogen bonding (H-bonding) and π interactions. Ionic interactions are typically strong and involve attraction for cation-anion and repulsion for cation-cation and anion-anion. Types of H-bonds that are common in ILs include traditional H-bonds from ammonium (N-H) and hydroxyl (O-H) donors, as well as interactions where ring C-H, alkyl C-H or N-H act as donor groups and anions as acceptors.[26] Although dispersion interactions are typically weaker than ionic or H-bonding, they are common for ILs that contains alkyl chains. These interactions can be substantial in large molecules that contain many atoms and can be competitive with strong ionic interactions. [26]

Imidazolium cations are aromatic and thus can form π–π stacking among themselves or with other aromatic molecules. The $C^2$-H attached to the imidazolium cation can also form H-bonding with alkyl sulfate anions (**Figure 3**).[26] When multiple types of interactions which are comparable in strength are present, it could be difficult to identify specific interactions. An example would be the competing π–π and H-bonding interactions which were uncovered in this study.[26]

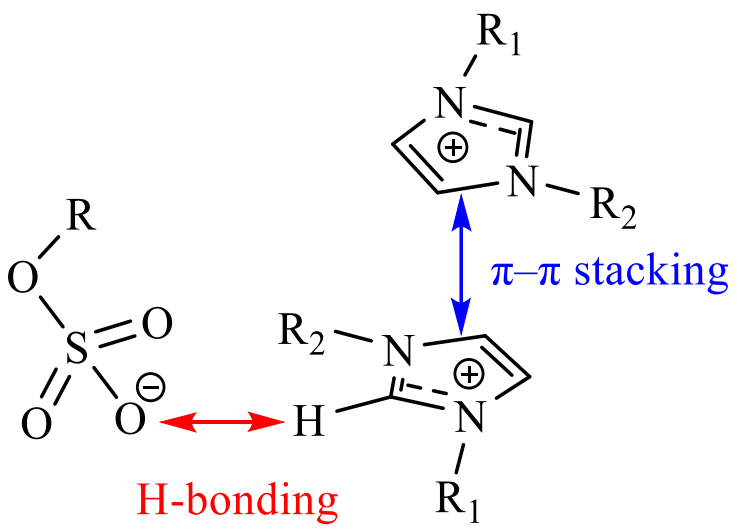

Figure 3: Interactions between imidazolium cations and alkyl sulfate anions.

### *Density functional theory (DFT)*

*Ab initio* quantum chemical (QC) calculations offer a way to investigate the chemical properties of solutes and ILs that can be difficult or time-consuming to examine experimentally.[27] Through *QC* methods, the structural, electronic, vibrational and NMR properties can be obtained. Gas-phase (GP) calculations of isolated molecules and clusters can also generate accurate association energies, electron density maps and electrostatic potentials (ESPs).[28]

Density functional theory (DFT) is a computational quantum mechanical modelling method often used to investigate the electronic structure of many-body systems in condensed phases. The properties of complex systems are determined using functionals of the spatially dependent electron density. DFT methods require a basis set specification. A basis set is a set of wave functions that describes the shape of atomic orbitals (AOs). The accuracy of the approximations used in DFT calculations are directly related to the basis sets used, while computational cost shows an inverse relation.[29] Hence, the choice of basis sets often faces a trade-off between accuracy of results and CPU time.

The popular DFT method B3LYP uses a hybrid functional consisting of the Hartree-Fock functional, Lee-Yang-Parr (LYP) correlation functional and the Becke 88 exchange functional.[30] Previous studies have found that DFT to be a suitable compromise between accuracy and computational cost for carrying out QC calculations on ILs.[31] Nevertheless, the applicability of DFT models depends on the nature of the system in question, and is influenced by molecular size, accuracy required, and types of interactions involved.[27 32]

Although other higher-level methods (such as the GGA+D3, SCS-IL-MP2, DFTB3-D methods)[33 34] could yield higher-quality results, only small molecules can be treated with high levels of theory.[27] This is due to the substantial increase in computational resource associated with using improved methods or examining larger molecules or ions, such as those found in ILs.[27] For the calculations similar to those con ducted in this study, previous work have found B3LYP/6-311+G(d,p) sufficient to produce reasonably accurate results.[26]Although the Pople basis set [6-311+G(d,p)] used here are less robust than other higher-level ones (such as the Dunning-Huzinaga basis sets[35]), the cost is substantially lowered both in terms of computational time and resource (cpu, memory and disk).[36]

Dispersion is an important consideration for ILs containing highly polarisable species that can undergo non-covalent associations; weak H-bonding, π-stacking and or anion–π interactions.[27] Dispersion effects can be significant, changing the relative geometry and orientation of the constituent ions within an IL.[37] For more accurate results, geometries must be computed with a dispersion-corrected method,[38] which is particularly important for the conformers of imidazolium-based ILs.[39] As such, dispersion considerations were taken into consideration in this study using Grimmes D3 dispersion correction with the damping function of Becke and Johnson (GD3-BJ).[40]

### *Implicit and Explicit Solvent Models*

Solvation effects can be investigated using implicit or explicit solvent models.[41] The latter uses a large number of explicitly modelled solvent molecules to surround the solutes,[42] while in implicit models the solvents are modelled as a continuous, homogeneous and isotropic medium.[43] Explicit models offer a very high level of details in modelling the effects of the actual solvent molecules within the solvation shell, but it is more complex and computationally expensive than implicit solvation models.[44] In contrast, implicit continuum solvation models are computationally less expensive, the explicit solvent is replaced by an averaged dielectric continuum.

QC calculations can employ different implicit solvation models, such as the polarizable continuum models (PCM), Solvation Models (SMx), and the Solvation Model based on Density (SMD) to represent solvation effects.[45] These models surround a molecule with a solvent cavity placing charges (modulated by the dielectric constant, ε, of the solvent) on the surface of the cavity in order to mimic the charge stabilisation of the solvent environment.[46] An example of the solvent cavity is illustrated in **Figure 4**. Inside the cavity the molecule still experiences a vacuum dielectric, so qualitatively the molecule is surrounded by the effect of the solvent environment created by charges at the cavity boundary, not by the explicit solvent molecules.[27]

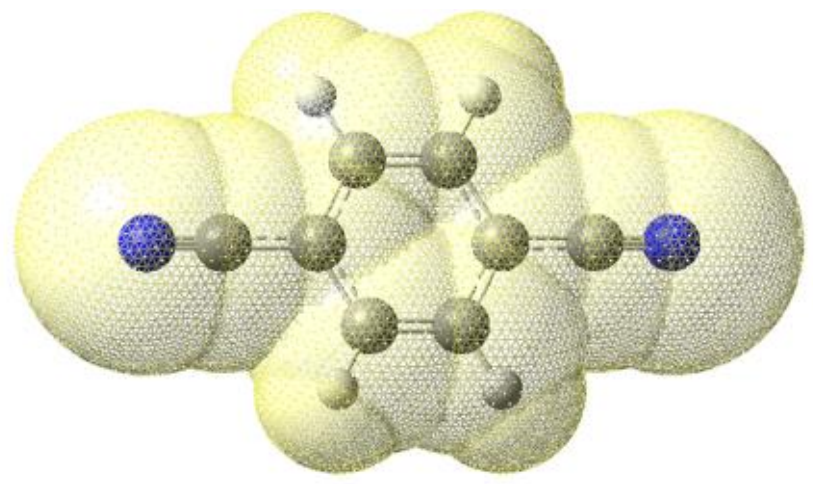

Figure 4: Solvated terephthalonitrile in the standard SMD solvent cavity of the IL, 1-Ethyl-3-methylimidazolium methyl sulfate

Using the SMD descriptors for ILs, the effects of a generalised solvent environment on the solvation can be explored.[27] SMD method computes solute–solvent interactions, but also includes terms due to cavitation, dispersion, repulsion and structure effects.[46] Thus, the SMD method requires the following additional solvent data: the index of refraction (n), the macroscopic surface tension (γ), the Abraham H-bond acidity (α), the Abraham H-bond basicity (β), the carbon aromaticity (φ) and the electronegative halogenicity (ψ).[46 47]

There are three types of SMD models commonly used. The standard SMD model uses all the parameters from the particular IL (or a very similar one) as the solvent environment. The generic ionic liquid model (SMD-GIL) uses average values of the above parameters obtained from a series of (slightly) structurally different imidazolium based ILs, except φ and ψ, which are calculated for the specific ILs used.[47] The partial generic parameters model (SMD-PGP) uses actual parameter values of the IL where possible, and average values where the actual values are not known.[47] In this project, all three solvation models were used, and the solvation energies calculated were compared to examine the difference in solvation effects introduced by different solvent properties.

The accuracy of results from implicit models could be limited by the omission of specific solvent-solute interactions (such as H-bonding).[48] These effects are particularly important in the regions close to the solute where the solvent has properties different from the bulk (first solvation shell).[49] To account for these, explicit models with a supermolecular or cluster-based approach are often used.[42]

The cluster approach used in this study involves only one solute molecule and ion(s) from a single ionic liquid ion pair (IL-IP). Although studies with a very low number of solvent and solute molecules might result in differences when compared to studies employing more molecules, the latter involves a significant computational cost due to the large number of solvent molecules required to model the bulk solution.[44] Moreover, keeping the solvation system simple allows the interactions to be understood more easily. When many solvent molecules are present, it could prove difficult to identify key geometries and dominant interactions. Hence, it is advisable to first investigate these microsolvated clusters involving one solute molecule and one IL-IP without a surrounding continuum to identify the most favourable interactions between the solute and IL solvent used.[50]

As both implicit and explicit methods have their strengths and weaknesses, each of these methods may be the preferred compromise of efficiency and expected reliability depending on particular problems and circumstances.[41] More recent work in computational modelling has developed a mixed method combining the two solvent models which could prove effective in modelling solvation.[42] This method constitutes the first solvation shell with the solute molecule and some number of explicit solvent molecules, while representing the remaining bulk solvent with a surrounding continuum.[49] This strategy is formally known as the discrete-continuum or semicontinuum solvation model[51] and has been previously used in several QC studies.[52] [53] [54]

In an attempt to offer a better description of solvation in ILs, the final part of this study attempts to investigate a selected optimised geometry employing a semicontinuum model. Although there are issues involving how to determine the number and orientation of non-bulk solvent molecules to use,[27] valuable information about solvation can be learnt from the semicontinuum calculations with one or two explicit solvent molecules.[42] Future works could involve more extensive studies to develop better hybrid models for this particular solvent-solute system.

## Objectives:

This research project involves the study of the intermolecular interactions between two structurally similar organic analytes, 1,4-diethynylbenzene (BzCC) and terephthalonitrile (BzCN), with the IL 1-Ethyl-3-methylimidazolium methyl sulfate, $[C_2C_1Im][MeSO_4]$, as shown in **Figure 3** below:

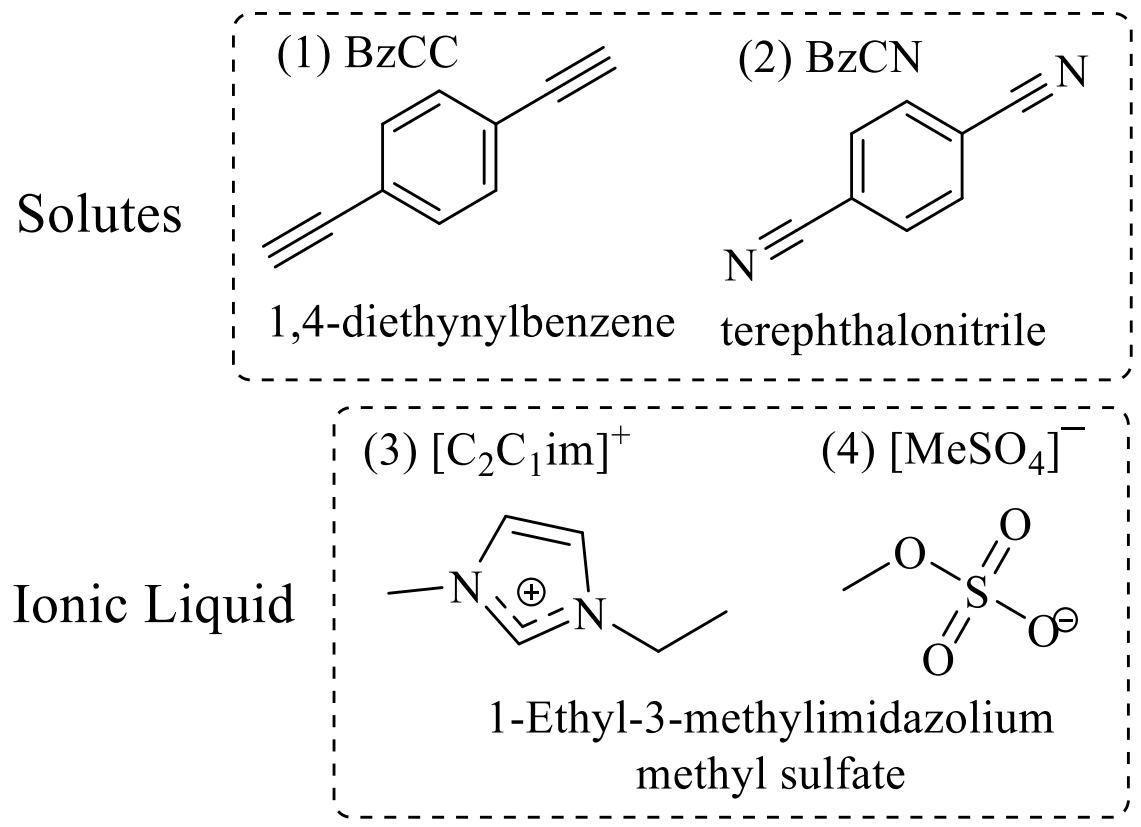

Figure 5: Solute and ionic liquid solvent molecules used in the study.

Gas-phase optimisation of the isolated geometries was performed, and the ESPs and ESP charges of the solutes and IL ions were visualised to predict the possible sites for non-covalent interactions. The conformational space of the ions interacting with each solute was then investigated. The stable ion-solute geometries and dominant interactions were then compared. Finally, sensible geometries of the IL-IP and the IL-IP with BzCC were constructed and optimised to understand the interactions between them, as well as to outline a further development of the project.

The solvation energies of each solute or ion in the ILs were then be investigated with the three SMD models (standard SMD, SMD-GIL, and SMD-PGP). The values were compared to determine the effects on solvation and molecular structure and geometry. The optimised IL-IP with BzCC geometries was further optimised with standard SMD. Possible further work includes conducting more in-depth calculations to understand the origins of the interactions and assess their relative strength.

# Methods

DFT calculations have been performed using Gaussian 09 (version D.01)[55] on the Imperial College High Performance Computing service. The initial geometries used in the calculations were constructed in GaussView (6.0.16) and optimized through employing the B3LYP hybrid functional and the 6-311+G(d,p) basis set.[56] To account for dispersion effects, GD3-BJ was used in every calculation.[40] The convergence criteria of $10^{-9}$ on the root-mean-square density matrix, and $10^{-7}$ on the energy matrix have been applied. A numerical integration grid of 99 radial shells and 590 angular points per shell were requested. The optimised structures have been confirmed as minima through checking for convergence and negative frequencies (with a cut-off of -20 $cm^{-1}$ to accommodate possible numerical inaccuracies during the calculations).

The isolated solutes and ions were investigated first in GP. The ESPs of the optimised solutes and IL ions were computed and used to identify potential sites for intermolecular interactions, as well as to guide the positioning of solute and ions in building clusters for subsequent calculations. This was achieved through identifying and matching the most positive and negative ESP positions so as to maximise attractive non-covalent interactions.

Guided by the ESPs, the supramolecular clusters consisting of the solute and either cation or anion of the IL were investigated. The input cluster structures were created by placing the molecules and ions around each other in a sensible manner to maximise attractive electrostatic, H-bonding, and π–π interactions. The resulting conformations were examined, and the association energy ($E_a$) of each geometry was calculated from the energy of the ion-solute cluster ($E_{cluster}$) and the sum of the total energies of the isolated components ($E_{components}$):

$$E_a = E_{cluster} - E_{components} \qquad (1)$$

Implicit solvation effects were then investigated using the different SMD models. Each of the optimised structures of the solutes and IL ions in GP were re-optimised with the SMD parameters (**Table 1**) of each SMD models: standard SMD, SMD-GIL, and SMD-PGP respectively.

The GP results were compared with the SMD results, and the solvation energies ($E_{solv}$) were derived from the energy difference between the optimised geometries in GP ($E_{GP}$) and in SMD ($E_{SMD}$ ) environment:

$$E_{solv} = E_{SMD} - E_{GP} \qquad (2)$$

The solvation energies obtained were used to derive a preliminary understanding of whether strong intermolecular interactions between the solutes and the ILs are expected.

Table 1: SMD parameters for the three SMD models used in the study, calculated from **Appendix.1**.[57]

| SMD parameters | Standard SMD | SMD-GIL[46] | SMD-PGP |
|---|---|---|---|
| Dielectric constant | 35[58] | 11.50 | 11.50 |
| Index of refraction, squared | 2.1730[59] | 2.0449 | 2.1730[59] |
| Macroscopic surface tension /cal $mol^{-1}$ $Å^{-2}$ | 56.84[59] | 61.24 | 56.84[59] |
| Abraham H-bond acidity | 0.224[60] | 0.229 | 0.229[46] |
| Abraham H-bond basicity | 0.525[60] | 0.265 | 0.265[46] |
| Fraction of non-H atoms which are aromatic carbon atoms | 0.2142 | 0.2142 | 0.2142 |
| Fraction of non-H atoms which are electronegative halogen atoms | 0 | 0 | 0 |

# Results and Discussion

## *Isolated solutes in GP*

The ESPs (**Figure 6**) and ESP charges (**Appendix.2-3**) were analysed, and the potential intermolecular interaction sites were identified. Neutral BzCC and BzCN were thought to be able to interact via either π-π stacking or H-bonding. The $[C_2C_1Im]^+$ and $[MeSO_4]^-$ were found to be good H-bond donors via the $C^2$-H and a good H-bond acceptor via the oxygen atoms respectively.

The neutral BzCC has charges mostly evenly distributed, which is expected due to the lack of polar functional group present in the molecule. Minor positive charges are accumulated at both terminal carbon of the alkyne group (**Figure 6**). For BzCN, the most negative electrostatic potentials have been found close to the two terminal nitrogen atoms of the cyanide group, which are expected to interact with the positive partial charges on the $C^2$-H of the $[C_2C_1Im]^+$ cation. The $C^2$-H could also participate in H-bonding with the oxygen atoms in the $[MeSO_4]^-$ anion.[61] For the anion, the oxygen atoms are expected to interact either through electrostatic interactions or H-bonding with the cation or solutes. Based on the ESPs obtained, solute-ion cluster geometries with stable interactions were predicted to be sandwiched for BzCC and H-bonded T-shape for BzCN.

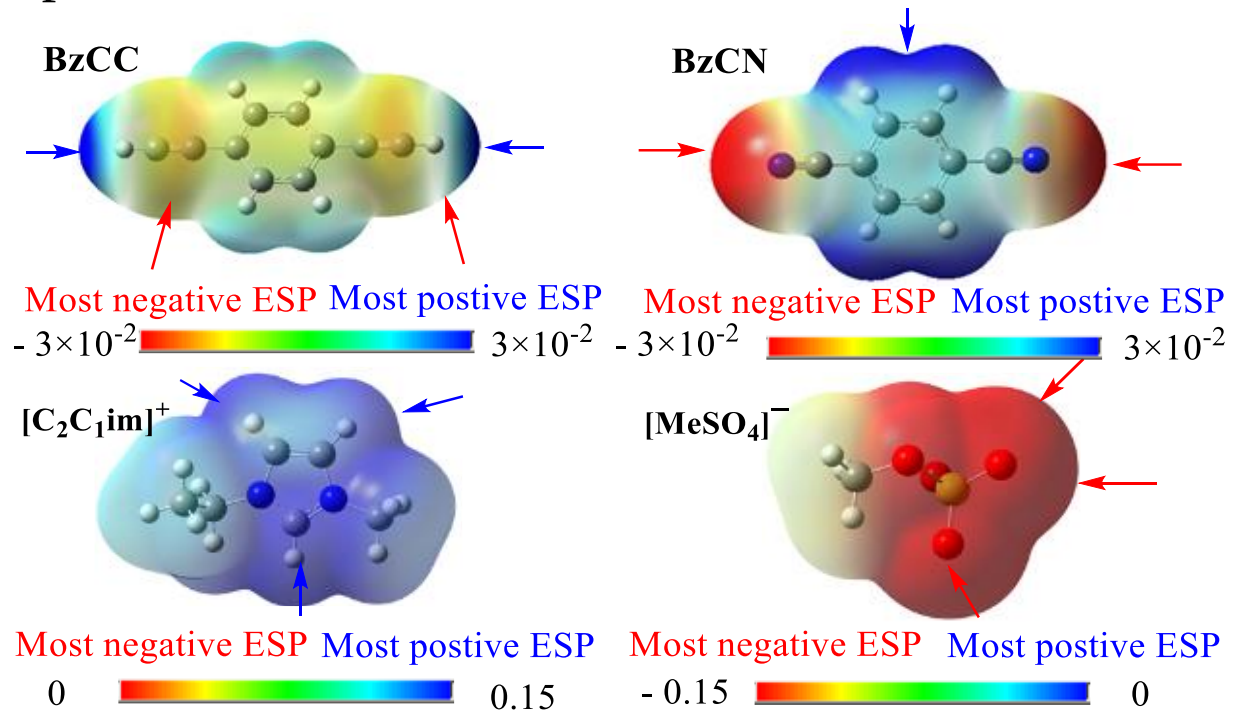

Figure 6: ESPs of solutes and IL ions.

These findings were then used to construct possible initial geometries of the solute-ion clusters (**Appendix.4**). For each solute and cation, variations of sandwiched, T-shaped, and parallel-displaced structures were constructed to maximise attractive non-covalent interactions between the solute and ion, allowing for ionic, H-bonding or π interactions to take place.[26] For each solute and anion, main initial geometries constructed included placing the $[MeSO_4]^-$ anion above the benzene ring, at the side or end of the solute molecules to maximise H-bonding.

## *Solute-ion clusters*

The geometries built for each solute-ion cluster were optimised in GP, and arranged in the order of association energy from most to least negative (**Appendix.5**). **Figure 7** shows the most stable geometry of each cluster. The geometries in **Figure 7** were mostly dominated by π-π stacking or H-bonding. Selected geometries in **Appendix.5** not dominated by H-bonding and their orbital interactions were highlighted in **Appendix.6**.

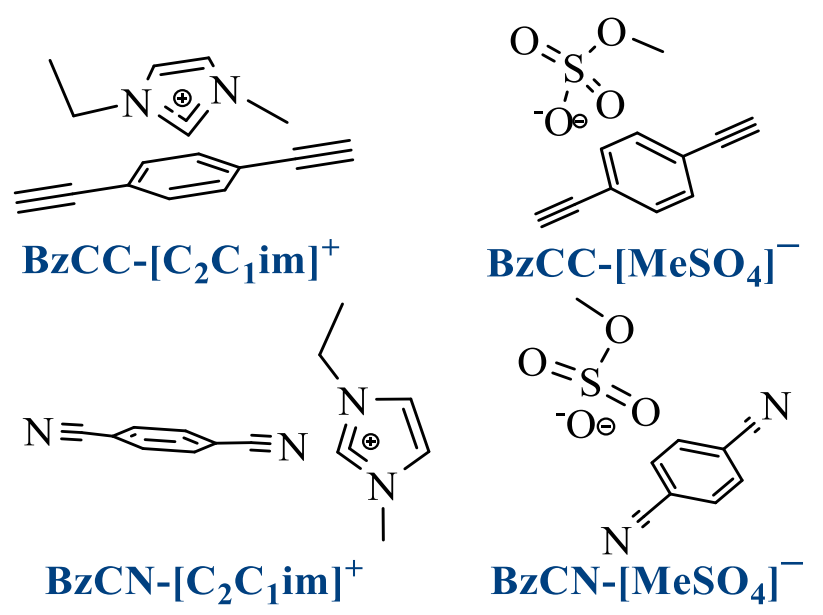

Figure 7: Most stable solute-ion geometries of each cluster.

The association energies for different initial solute-ion clusters (**Table 2**) were able to give an indication of how favourable the interactions are between the solute and ions in the optimised conformations. Optimised structures found to have less than 5 kJ mol$^{-1}$ energy difference could potentially be duplicate structures or have slightly different torsional angles.

Table 2: Association energies of solute-ion clusters.

| clusters | $E_a$ / kJ mol $^{-1}$ | | | |
|---|---|---|---|---|
| cluster No. | (1) | (2) | (3) | (4) |
| BzCC + $[C_2C_1im]^+$ | -61 | -59 | -55 | -50 |
| BzCC + $[MeSO4]^-$ | -40 | -47 | -43 | -41 |
| BzCN + $[C_2C_1im]^+$ | -52 | -53 | -45 | -36 |
| BzCN + $[MeSO4]^-$ | -84 | -83 | -82 | -68 |

The geometries for BzCC and the cation appeared to be mostly dominated by π-π stacking, where the benzene and imidazolium rings in the solute and ions were in parallel arrangements. The π-π stacking distance, 3.53 Å (measured as the distance between the two ring centroids, **Appendix.7**) was found to be comparable to the distance at which such π-π stacking interactions are significant[13] (3.656 Å for benzene). [62] This suggests that the π-interactions helped to lower the energy of the solute-ion cluster.

For BzCN and cation, the stable geometries showed H-bonding between the cyanide nitrogen and $C^2$-H of the cation, instead of a parallel π-stacking structure. This resulted in a H-bonded T-shaped structure, with the most stable geometry identified having the ethyl group on the imidazolium ring closer to the solute than the methyl group (**Figure 7**).

For both BzCC and BzCN, the most stable geometry with the anion had the anion located at side of the solute, next to the benzene ring. These side-on geometries could be explained by the double H-bonding between the two ring protons and the anion oxygens. For BzCC, this stable geometry differed from what was expected from solely analysing the charge distribution, where the most negative end of the anion was expected to interact with the most positive end (alkyne) of the solute in an end-on in a linear manner. While for BzCN, this geometry was consistent with the expectation from their ESP analysis, where the most negative end of the anion interacts with the most positive end of the solute (ring hydrogen atoms). Notably, the association energies with BzCN was almost twice as negative of that of BzCC, showing that despite having the same number of H-bonding, the strength of which was influenced by greater charge distribution in BzCN.

## *SMD optimisation and solvation energy*

Table 3: Solvation Energies calculated from the three SMD solvation models. $E_{solv} = E_{SMD} - E_{GP}$, ΔEs are calculates with respect to standard SMD.

| Solutes and IL ions | solvation energy, $E_{solv}$ / kJ mol $^{-1}$ | | | | |
|---|---|---|---|---|---|
| | Standard SMD | SMD GIL | GIL ΔE | SMD PGP | PGP ΔE |
| BzCC | -21 | -21 | 0 | -23 | -2 |
| BzCN | -40 | -36 | +4 | -38 | +2 |
| $[C_2C_1im]^+$ | -239 | -224 | +15 | -226 | +13 |
| $[MeSO4]^-$ | -244 | -222 | +2 | -223 | +1 |

Three SMD (standard SMD, SMD-GIL, and SMD-PGP) models were employed to optimise the isolated analytes and $E_{solv}$ was calculated for each model **(Table 3)**. Considering an error margin of roughly 5 kJ mol$^{-1}$ from the functionals and basis sets used, $E_{solv}$ of each analyte were fairly constant across the three different SMD models used except for $[C_2C_1im]^+$, where the differences between $E_{solv}$ of the standard SMD and of SMD-GIL or SMD-PGP were more significant. However, the differences in $E_{solv}$ between different SMD models did not seem to affect the optimised geometries.

The geometries calculated suggested similar optimised structure in all three SMD environments, as using VMD to overlay the structures produced identical overlapping geometries, even for $[C_2C_1im]^+$ which had the greatest ΔE between the different SMD models. Closer look at the selected bond lengths and angles for standard SMD and SMD-GIL models of each solute or ion (**Appendix.8**) corroborated with the above findings. The bond length and angles of BzCC, BzCN and $[C_2C_1im]^+$ are identical to the accuracy given. Small differences in the geometries between SMD-GIL and standard SMD for $[MeSO_4]^-$ were found, mainly in the bond angles (116.2° vs 116.0° and 105.7° vs 105.6°). However, the calculations might not produce accuracy up to one decimal, so the geometries could potentially be identical. These geometrical differences could be due to the charged $[MeSO_4]^-$ ions having stronger interactions with the IL solvent, hence would be better solvated.

The results also suggested that the dielectric constant play a greater role compared to other SMD parameters in defining the solvation environment. By comparing the parameters of standard SMD and SMD-GIL or SMD-PGP (which have the most significant difference in dielectric constant values), the ΔE between the standard SMD and SMD-GIL tend to be larger than that between SMD-GIL and SMD-PGP. Refractive index, macroscopic surface tension and Abraham H-bond basicity play a less prominent role in defining the solvent environment. This conclusion could be drawn by examining the $E_{solv}$ differences between SMD-GIL and SMD-PGP.

Standard SMD models were used to calculate the most stable geometry of each optimised solute-ion clusters in GP (**Figure 7**). The resultant $E_{solv}$ were shown in **Table 4**.

Table 4: $E_{solv}$ of the most stable solute-ion clusters geometires.

| Cluster | $E_{solv}$ / kJ mol$^{-1}$ |
|---|---|
| | Most stable geometry |
| BzCC + $[C_2C_1im]^+$ | -231 |
| BzCC + $[MeSO4]^-$ | -234 |
| BzCN + $[C_2C_1im]^+$ | -224 |
| BzCN + $[MeSO4]^-$ | -218 |

The $E_{solv}$ here were significantly larger than the $E_a$ in the previous section (**Table 2**) where individual solute and ion interacts explicitly in GP. This suggests the implicit solvation of the solute-ion cluster created significantly more stabilisation to the cluster in GP.

## *Ionic liquid ion pair with BzCC:*

With the above preliminary calculations of the isolated solutes, isolated IL ions and solute-ion clusters in GP and SMD, the project went on to optimise an IL-IP geometry. One possible geometry of the IL-IP with BzCC was also optimised in both GP and SMD.

IL-IPs are the simplest substructures of an IL to enable the study of cation-anion interactions. The IL-IP used in this study, $[C_2C_1Im][MeSO_4]$, was obtained through modifying the lowest energy optimised geometry of the structurally-similar IL-IP $[C_2C_1Im][EtSO_4]$ (obtained from Hunt Group database), by replacing the ethyl with methyl functional group. The resultant IL-IP geometry was then optimised in the GP with the anion above the imidazolium ring (**Figure 8**), indicating possible lone pair-π interactions, anion-π interactions, or H-bonding.

Other geometries of the IL-IP are possible as the optimised IL-IP geometry could be highly dependent on the initial geometry. Hence, future work could attempt to investigate different conformations of this IL-IP. Dimer structures could also be investigated to examine cation-cation and anion-anion interactions, as studies have found IL-IP dimers to be a more realistic model for bulk ILs than a single IL-IP.[44]

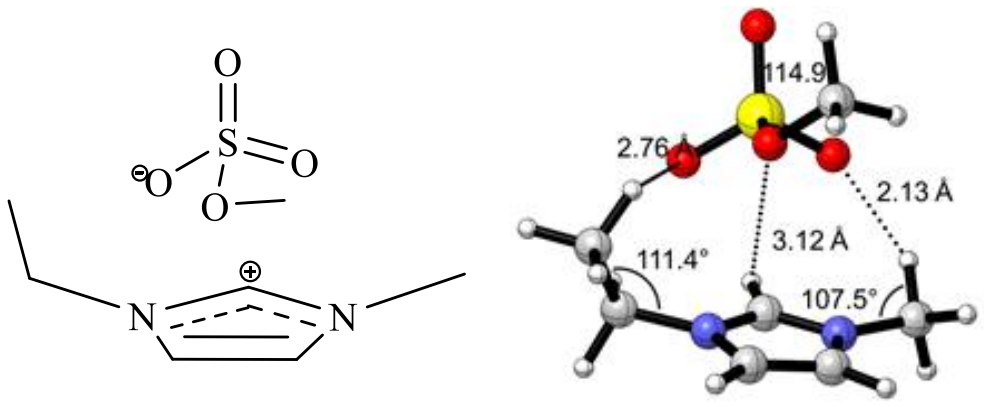

Figure 8: Optimised geometry of IL-IP of $[C_2C_1Im][MeSO_4]$ in GP.

Subsequently, the solute BzCC was introduced to the optimised IL-IP and further optimised in GP, **Figure 9**. The IL-IP geometry was largely retained and the distances between BzCC and the IL-IP were found to be significantly larger than those between the ions. This suggested the strong interactions between the IL-IP were not disrupted by introduction of BzCC. While there might be weak association between the ions and BzCC, such as the H-bonding between the oxygen atom and benzene-ring protons, these interactions tend to be weak.

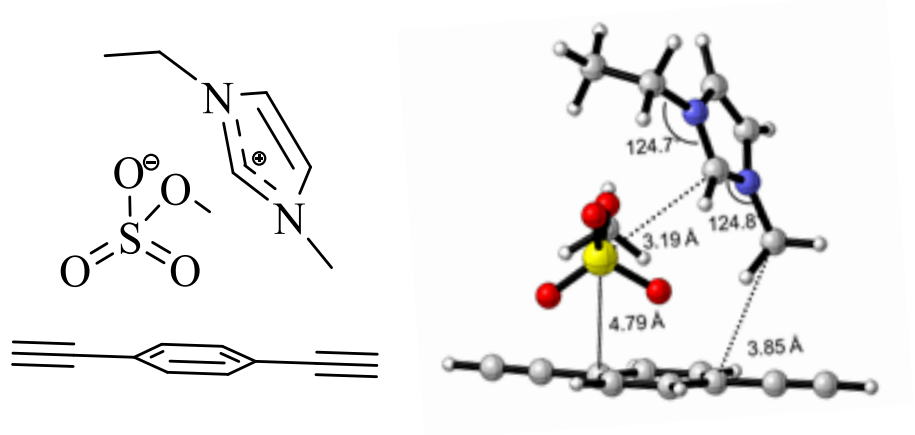

Figure 9: Optimised BzCC and $[C_2C_1Im][MeSO_4]$ IL-IP geometry in GP.

The optimised cluster geometry of BzCC and IL-IP were then calculated in the standard SMD environment (**Figure 10)** in attempt to construct a simple semicontinuum model. There were only minor changes to the selected interatomic distances and bond angles, but the association energy of the semicontinuum model was evidently more negative (**Table 5**), suggesting further stabilisation of the cluster compared to the explicit model only in GP. More calculations on different starting geometries of the BzCC and IL-IP cluster in GP and SMD could be conducted to prove that the above findings are consistent.

However, it is worth noting that the semicontinuum model adopted here is a very simplified one involving only one explicitly solvated IL-IP. Further work could aim to identify and develop better hybrid models for this particular solvent-solute system.

Table 5: Association energies of BzCC & $[C_2C_1Im][MeSO_4]$ IL-IP clusters in GP and standard SMD.

| $E_a$ / kJ mol$^{-1}$ | BzCC and IL-IP cluster |
|---|---|
| GP | *-432.66* |
| Standard SMD | *-560.56* |

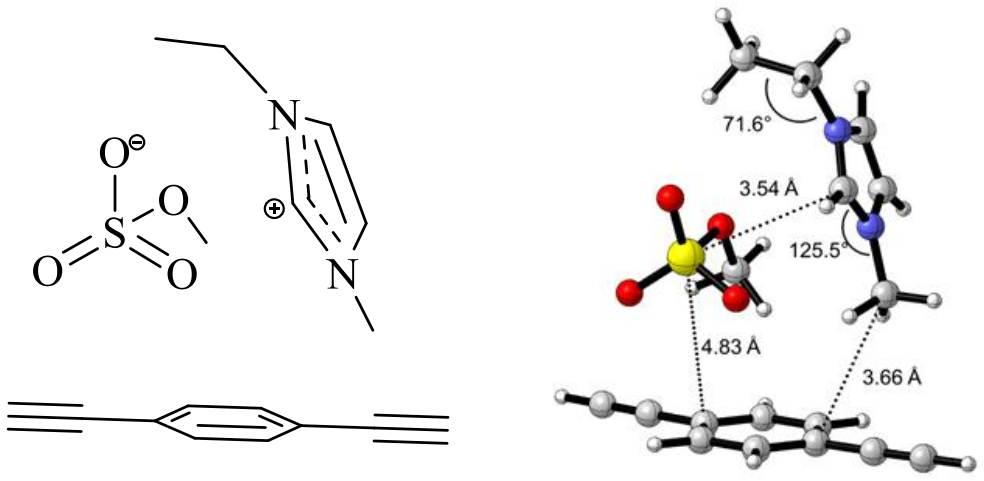

Figure 10: Optimised BzCC and $[C_2C_1Im][MeSO_4]$ IL-IP geometry in standard SMD environment.

## Conclusions

In this study, DFT methods have been successfully used to investigate the intermolecular interactions between two structurally similar solid organic analytes, 1,4-diethynylbenzene and terephthalonitrile, with the ionic liquid $[C_2C_1Im][MeSO_4]$. Through initial analysis of isolated solute and ions, potential intermolecular interaction sites have been identified and used for the construction of solute-ion clusters. The clusters were optimised in gas-phase and the stable geometries were compared. BzCC and BzCN were found to interact with the cation via π–π stacking and H-bonding respectively, and both solutes interact with the anion mainly through H-bonding. Future work could involve more in-depth calculations to understand the origins of these interactions and assess their relative strength, as well as comparison of the computational results with experimental values.

The individual solutes and ions were calculated in different SMD models (standard SMD, SMD-GIL and SMD-PGP) and the respective solvation energies and optimised geometries were compared. Selected solute-ion and a BzCC IL-IP clusters were then optimised in standard SMD in an attempt to construct a simplified semicontinuum model. Implicit solvation prove to offer greater stabilisation to the molecular clusters than the isolated clusters in gas phase alone for the particular case examined, despite causing very little change to the cluster geometries. Future work could involve developing better semicontinuum model or compare the computational results with experimental values.

# Appendix

## Appendix.1: Additional descriptions and calculations for the SMD parameters used.

**Table 6: Macroscopic solvent descriptors for 12 ILs used for SMD-GIL parameters.**[57]

| ionic liquid | $\varepsilon$ | $n$ | $\gamma$ | $\varphi$ | $\psi$ |
|---|---|---|---|---|---|
| [BMIM][$BF_4$] | 11.70[81] | 1.4215[86] | 67.07[87] | 0.2000 | 0.2667 |
| [BMIM][$NTf_2$] | 11.52[85] | 1.4271[87] | 53.97[87] | 0.1200 | 0.2400 |
| [BMIM][$PF_6$] | 11.40[81] | 1.4090[87] | 70.24[87] | 0.1765 | 0.3529 |
| [EMIM][$BF_4$] | 12.80[81] | 1.4098[82] | 78.30[83] | 0.2308 | 0.3077 |
| [EMIM][DCA] | 11.00[84] | 1.5329[82] | 66.07[82] | 0.2308 | 0.0000 |
| [EMIM][$NTf_2$] | 12.25[85] | 1.4225[86] | 56.13[83] | 0.1304 | 0.2609 |
| [EMIM][OTf] | 15.20[81] | | 56.42[83] | 0.1875 | 0.1875 |
| [HMIM][$NTf_2$] | 12.70[84] | 1.4295[86] | 50.38[83] | 0.1111 | 0.2222 |
| [HMIM][$PF_6$] | 8.90[81] | 1.4165[86] | 62.47[87] | 0.1579 | 0.3158 |
| [$M_3$BAm][$NTf_2$] | | | | 0.0000 | 0.2609 |
| [MBPy][$BF_4$] | | | | 0.3125 | 0.2500 |
| [OMIM][$PF_6$] | 7.50[88] | 1.4235[89] | 51.38[89] | 0.1429 | 0.2857 |

[a]All descriptors are dimensionless except $\gamma$, which has units of cal $mol^{-1}$ $Å^{-2}$. Superscript numbers to the right of values are literature citations, not uncertainties.

**Dielectric constant (eps)** for [$C_2C_1$Im][$MeSO_4$] = 35[58]

The dielectric constant of [$C_2C_1$Im][$EtSO_4$] was used for [$C_2C_1$Im][$MeSO_4$]. It is a reasonable value to use due to the high structure similarity of the two ILs.

**Kamlet-Taft vs Abraham H-bonding parameters ($\Sigma\alpha_2^H$, $\Sigma\beta_2^H$) for [$C_2C_1$Im][$MeSO_4$].**

$\Sigma\alpha_2^H$, $\Sigma\beta_2^H$ are calculated from Kamlet-Taft ($\alpha$, $\beta$) measurements:

$\alpha = 0.53$,[59] $b = 0.71$[59]

$$\Sigma\alpha_2^H = 0.4098\alpha + 0.0064 = 0.223594 = 0.224$$

$$\Sigma\beta_2^H = 0.6138\beta + 0.0890 = 0.524798 = 0.525$$

**CarbonAromaticity ($\varphi$)** for [$C_2C_1$Im][$MeSO_4$].

There are 14 non-H atoms, 3 are aromatic C atoms, $\varphi = 3/14=0.2142$

**ElectronegativeHalogenicity ($\psi$)** for [$C_2C_1$Im][$MeSO_4$].

There are no electronegative halogen atoms $\psi = 0.0$

## Appendix.2: Table of ESP charges and ESP of the GP analytes.

<table>
<tr><th>GP Analytes</th><th>ESP derived Charges</th><th>Electrostatic potential(ESP)</th></tr>
<tr><td>BzCC</td><td>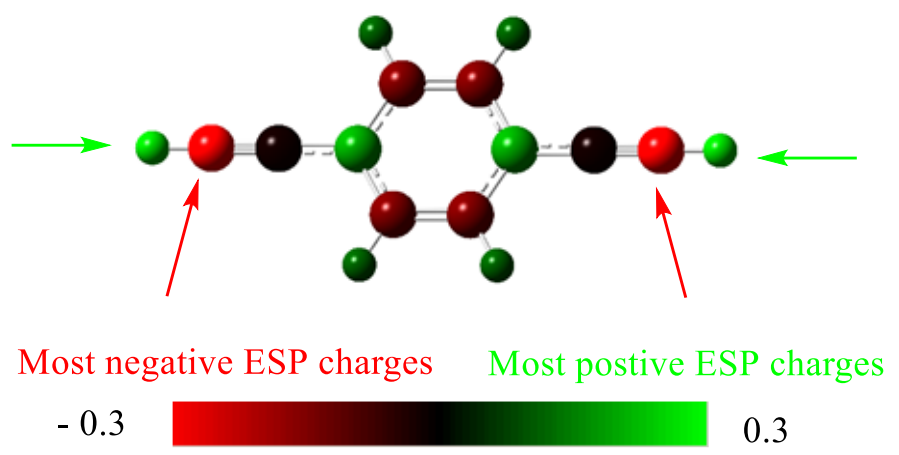
</td><td>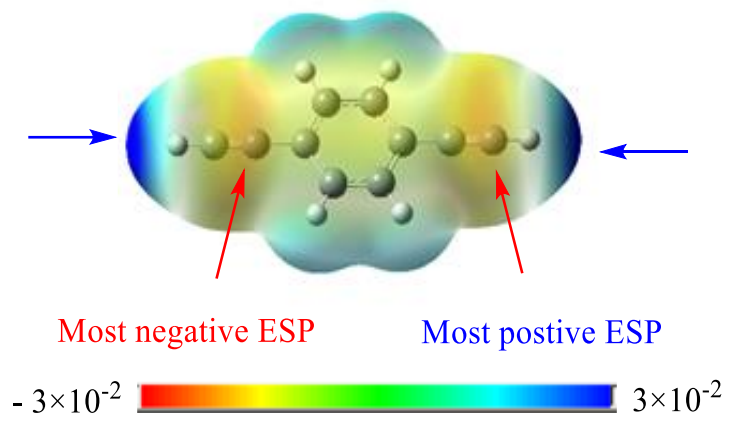
</td></tr>
<tr><td>BzCN</td><td>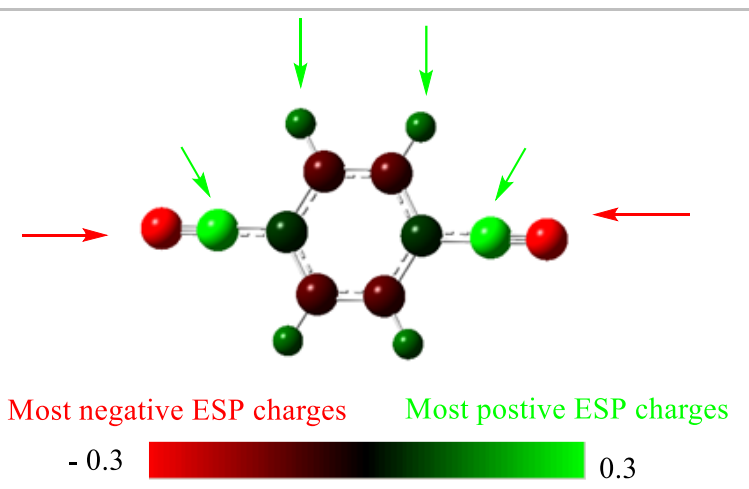
</td><td>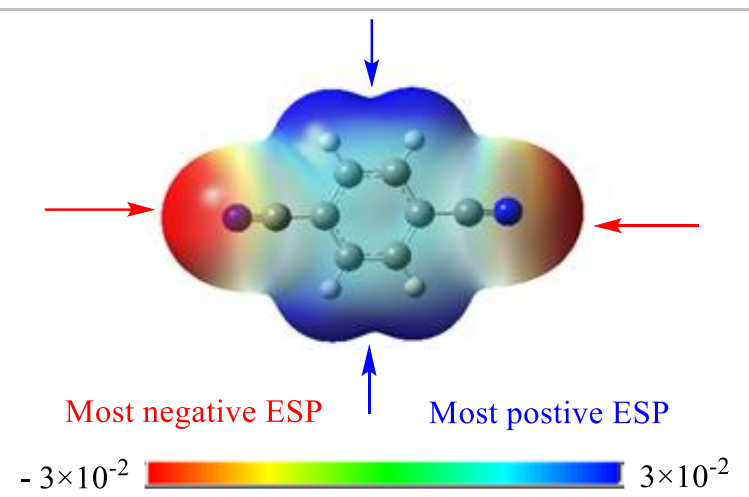
</td></tr>
<tr><td>$[C_2C_1m]^+$</td><td>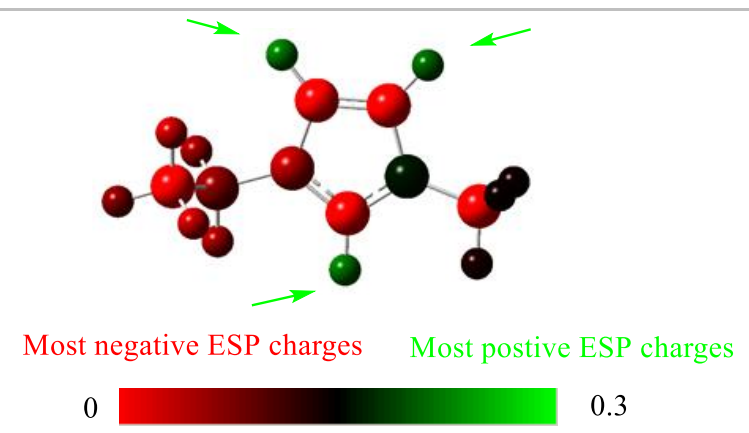
</td><td>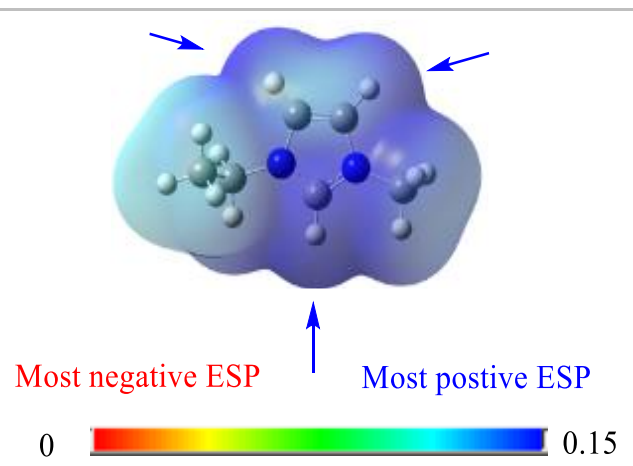
</td></tr>
<tr><td>$[MeSO_4]^-$</td><td>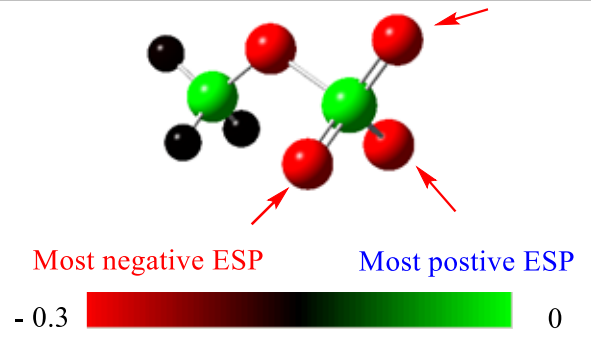
</td><td>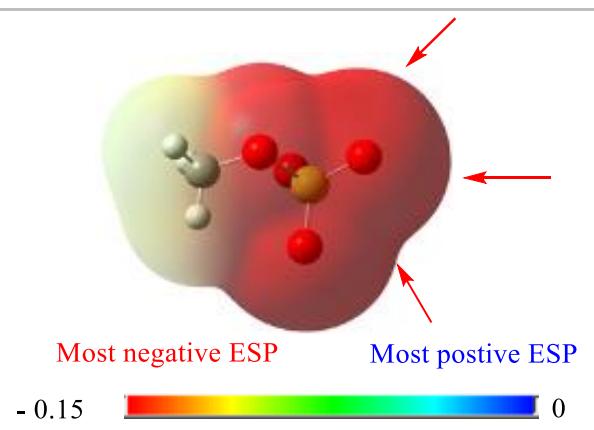
</td></tr>
</table>

## Appendix.3: Table of ESP charges labelled by atoms.

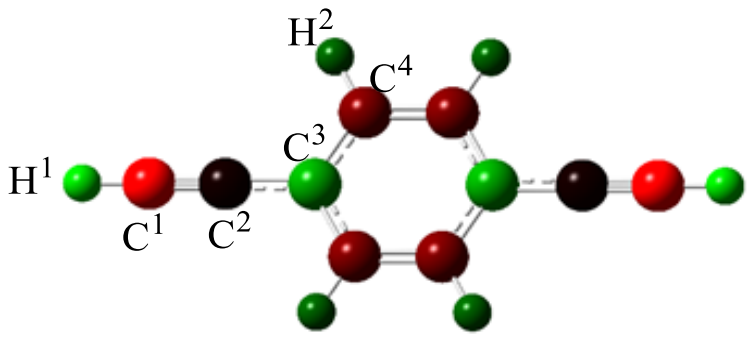

| | | |
|---|---|---|
| $H^1$: 0.237 | $C^1$: -0.335 | $C^3$:0.181 |
| $H^2$: 0.107 | $C^2$: -0.023 | $C^4$: -0.137 |

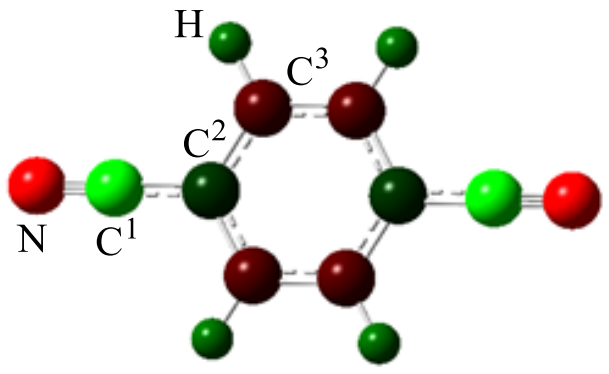

| | | |
|---|---|---|
| N: -0.451 | $C^1$:0.349 | $C^3$:-0.098 |
| H: 0.121 | $C^2$:0.057 | |

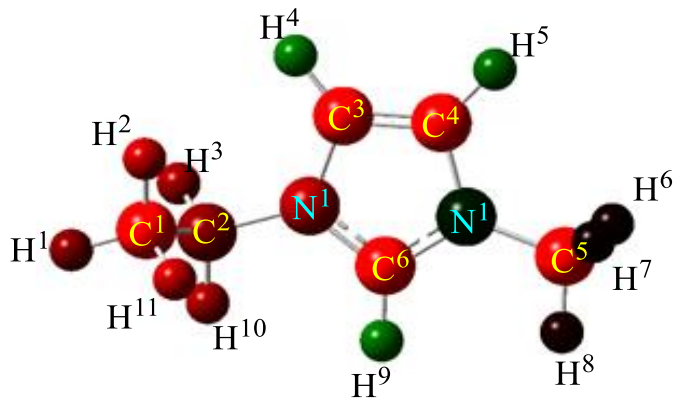

| | | | | |
|---|---|---|---|---|
| $H^1$: 0.080 | $H^5$: 0.204 | $H^9$: 0.211 | $C^2$: 0.075 | $C^6$: -0.067 |
| $H^2$: 0.045 | $H^6$:0.141 | $H^{10}$: 0.063 | $C^3$: -0.124 | $N^1$: 0.054 |
| $H^3$: 0.064 | $H^7$:0.144 | $H^{11}$: 0.044 | $C^4$: -0.134 | $N^1$:0.163 |
| $H^4$: 0.214 | $H^8$:0.136 | $C^1$: -0.088 | $C^5$: -0.223 | |

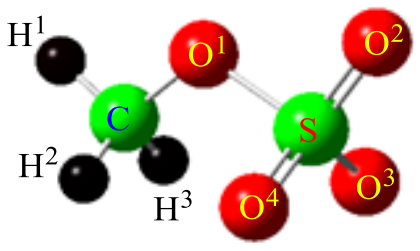

| | | |
|---|---|---|
| $H^1$: -0.018 | C: 0.192 | $O^2$: -0.700 |
| $H^2$: -0.002 | S: 1.418 | $O^3$: -0.700 |
| $H^3$: -0.002 | $O^1$: -0.494 | $O^4$: -0.700 |

## Appendix.4: Table of Starting geometries for solute-ion clusters.

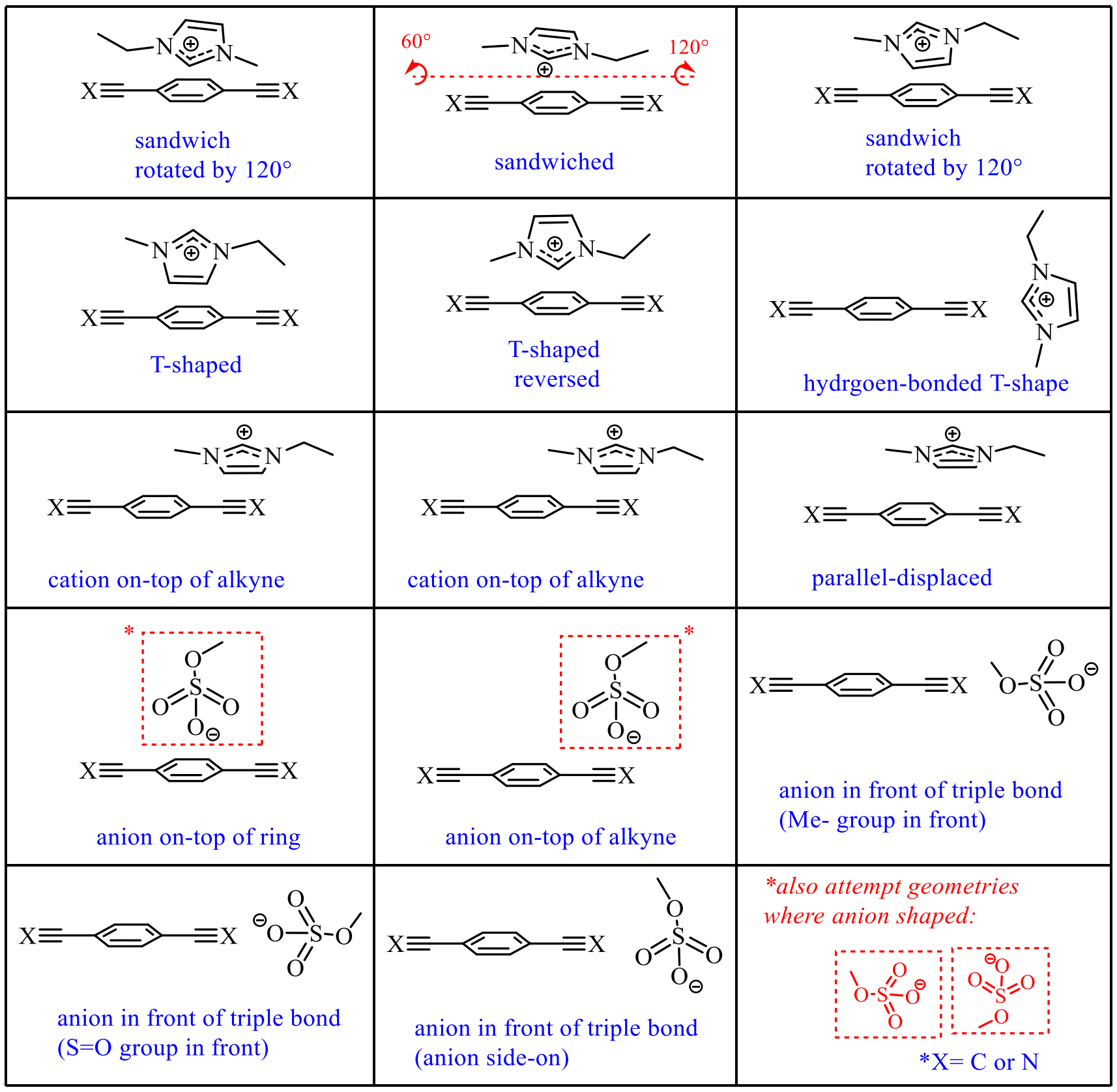

**Appendix.5: Table of optimised geometries for each solute-ion cluster arranged by their association energies (more most to least negative).**

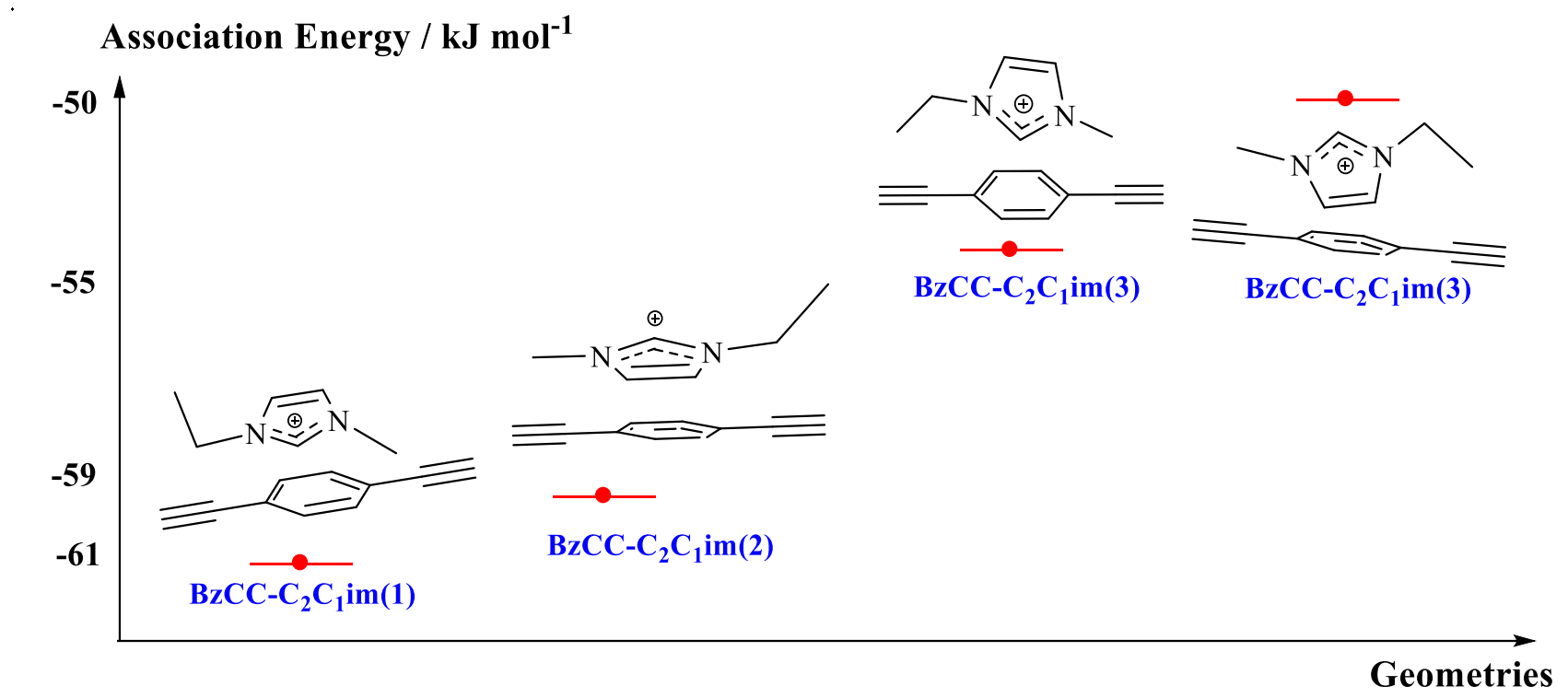

Figure 11: optimised geometries of BzCC + $[C_2C_1Im]^+$ cation.

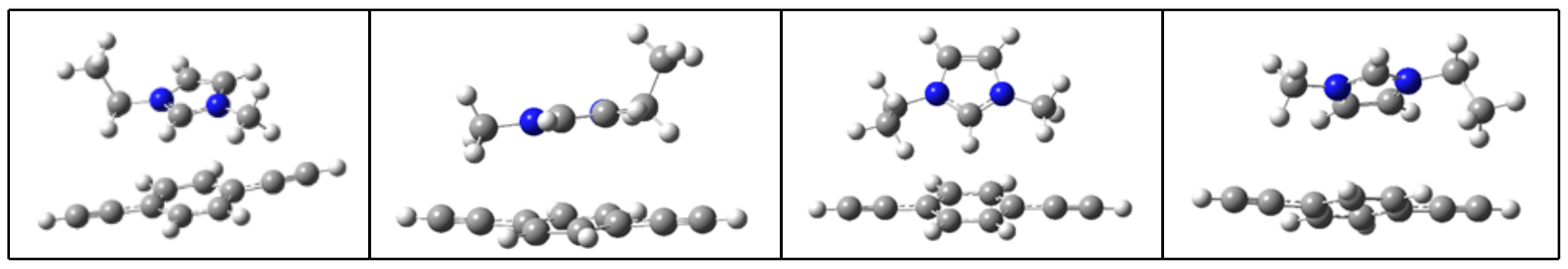

BzCC-$C_2C_1$im(1) BzCC-$C_2C_1$im(2) BzCC-$C_2C_1$im(3) BzCC-$C_2C_1$im(3)

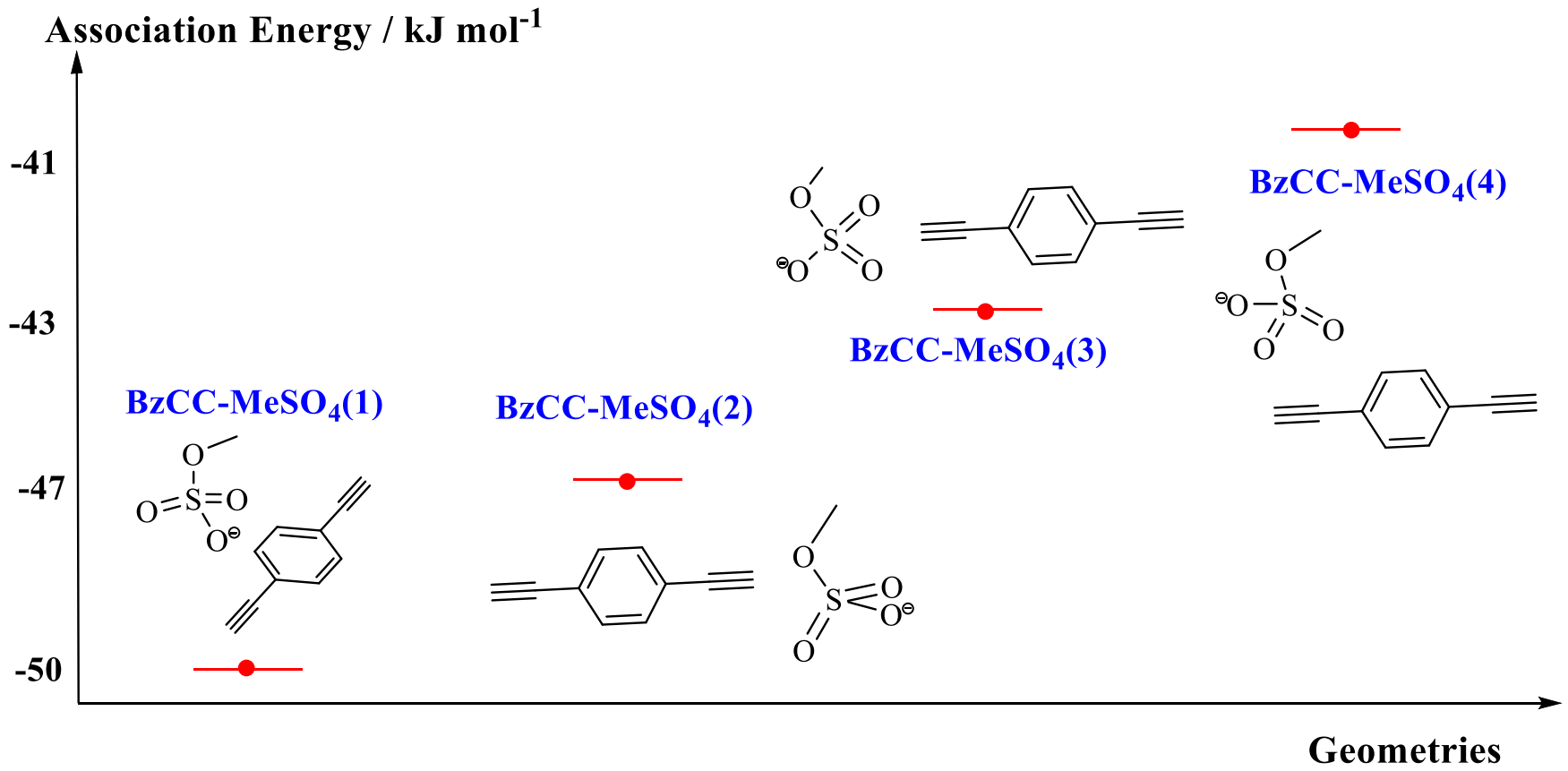

Figure 12: optimised geometries of BzCC + $[MeSO4]^-$ anion.

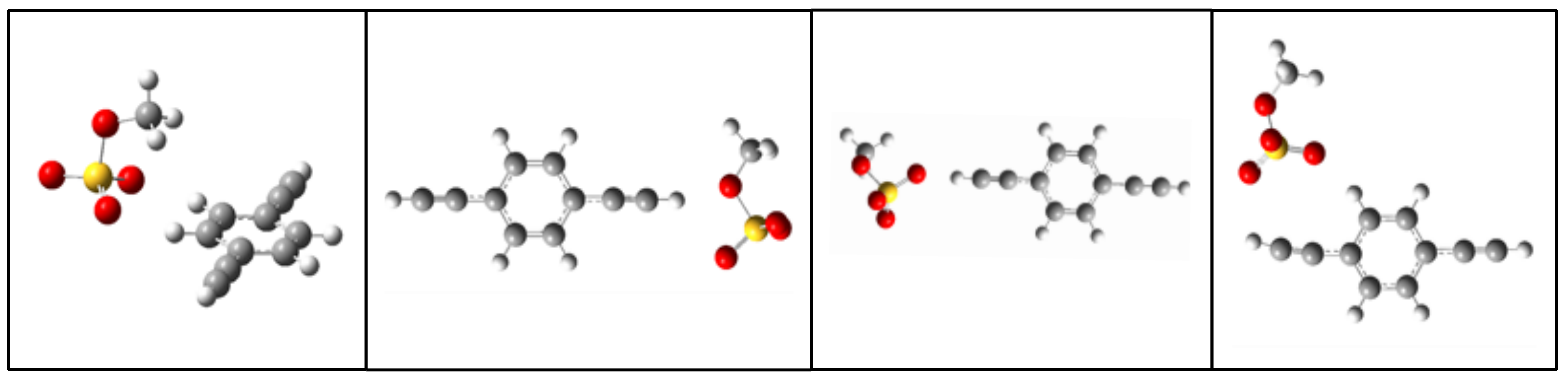

BzCC-$MeSO_4$(1) BzCC-$MeSO_4$(2) BzCC-$MeSO_4$(3) BzCC-$MeSO_4$(4)

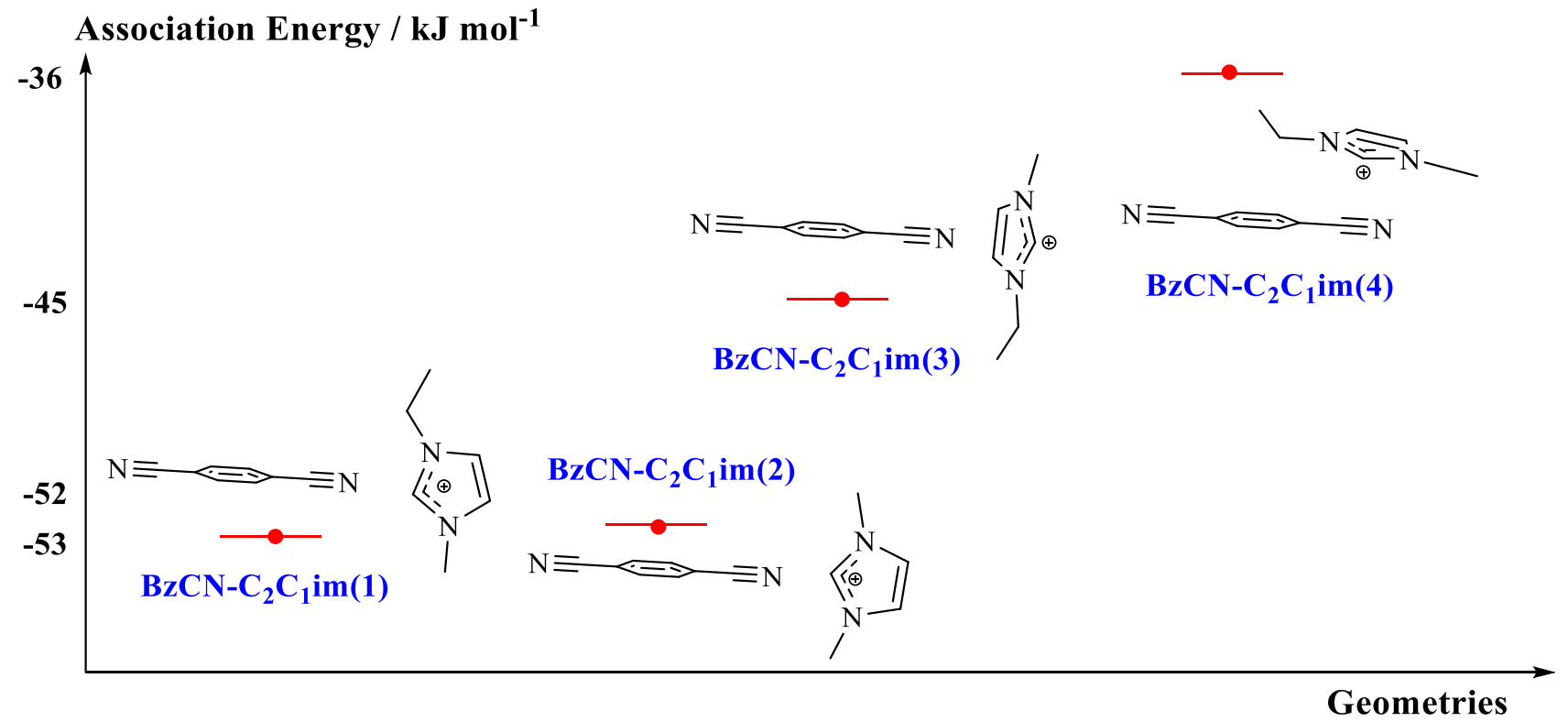

Figure 13: optimised geometries of BzCN + $[C_2C_1Im]^+$ cation.

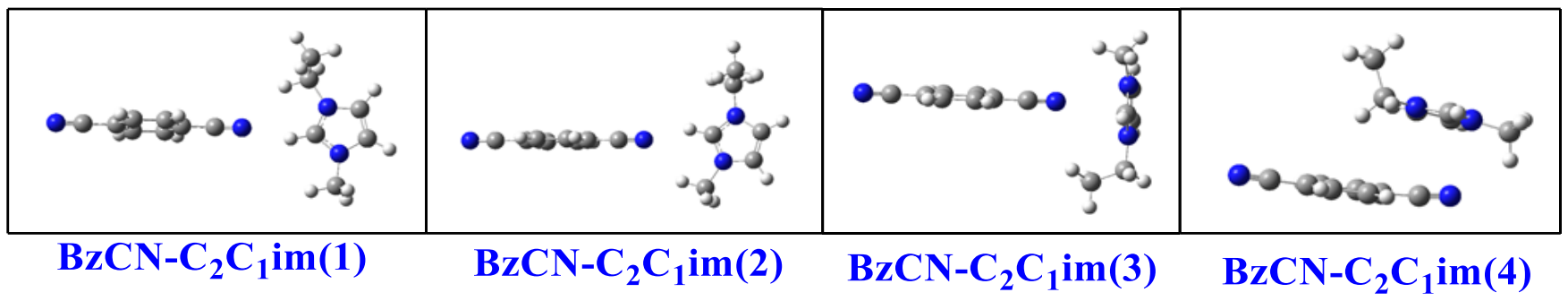

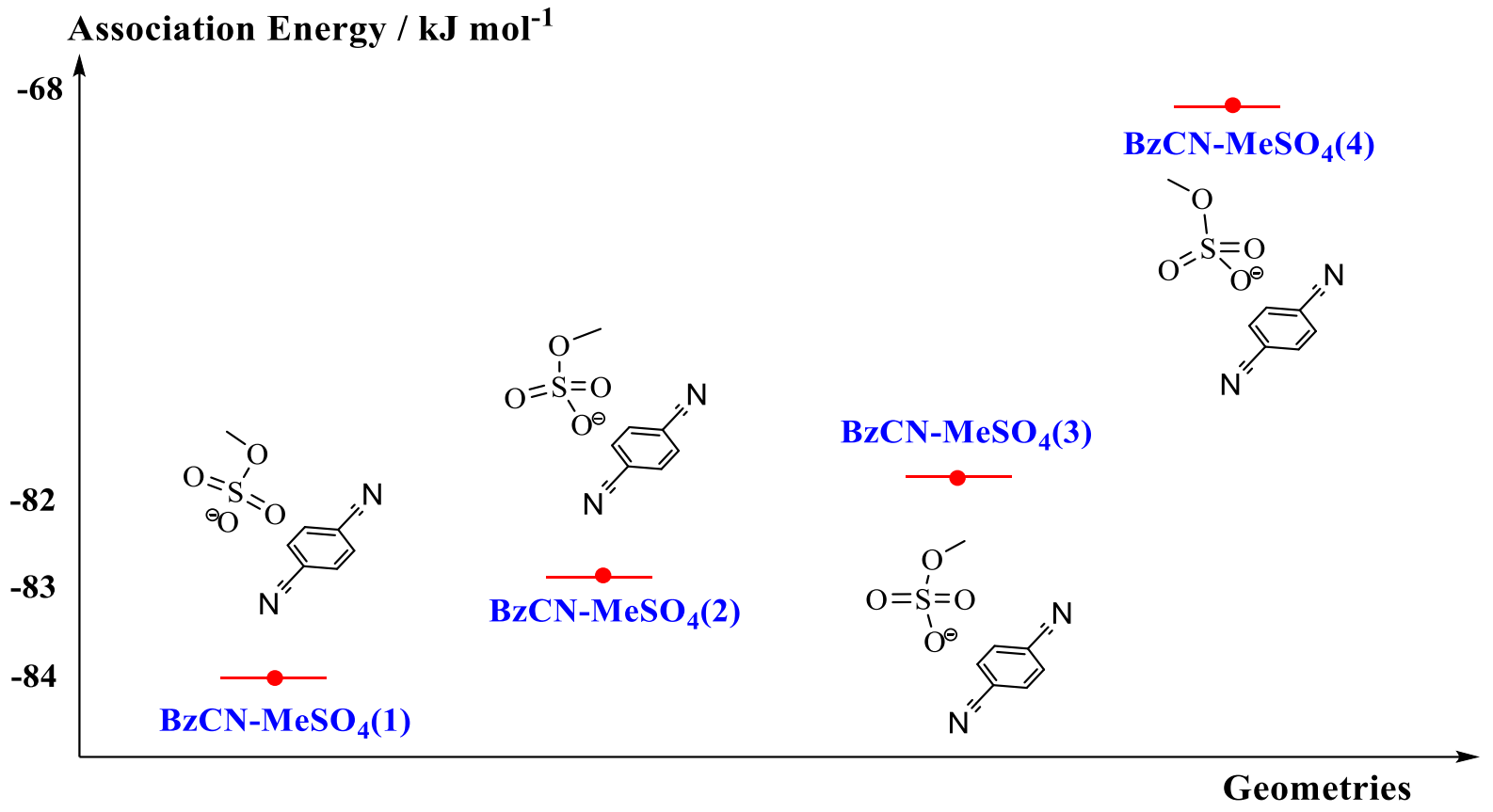

Figure 14: optimised geometries of BzCN +$[MeSO4]^-$ anion.

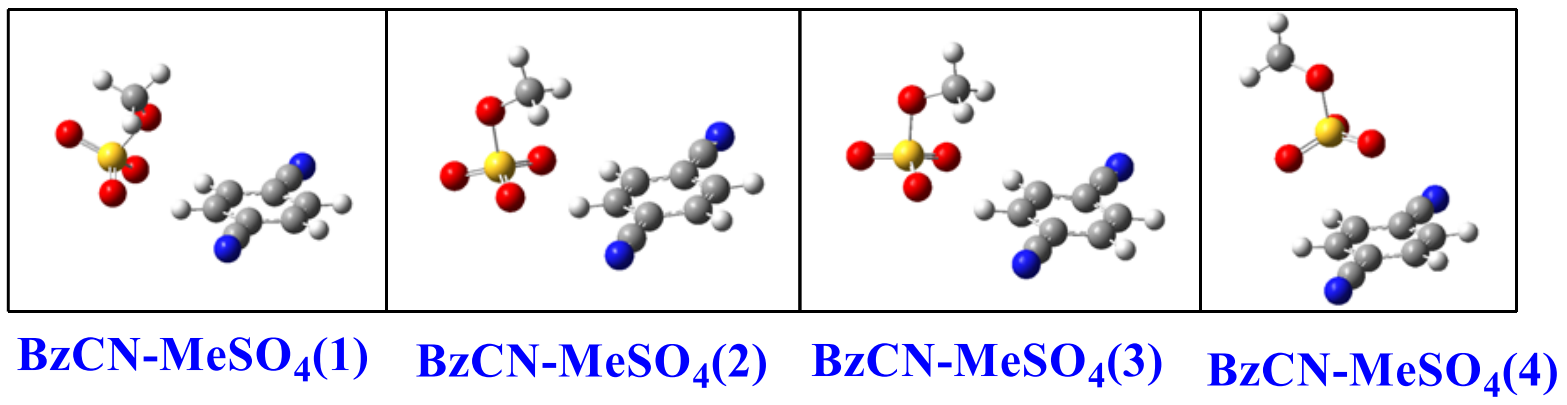

**Appendix.6: Bonding orbitals of interesting structures not dominated by hydrogen bonding BzCN-$[C_2C_1Im]^+$ (3) and BzCN-$[C_2C_1Im]^+$ (4).**

| | |
|---|---|
| **BzCN-$[C_2C_1Im]^+$ (3)**<br><br>Isovalue = 0.01 | 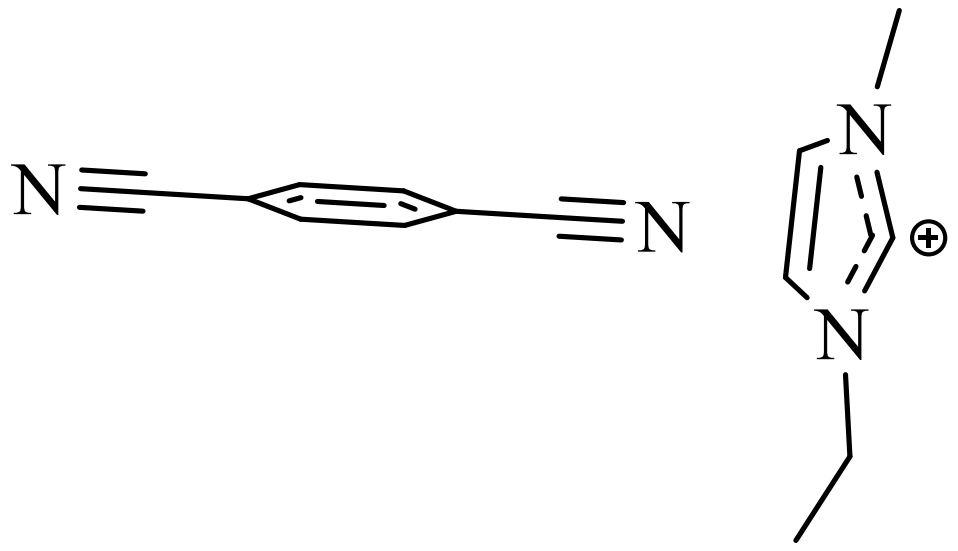  |
| **HOMO – 15**<br><br>Overlap of CN π electrons in BzCN with the π electron density of the $[C_2C_1Im]^+$ ring. | 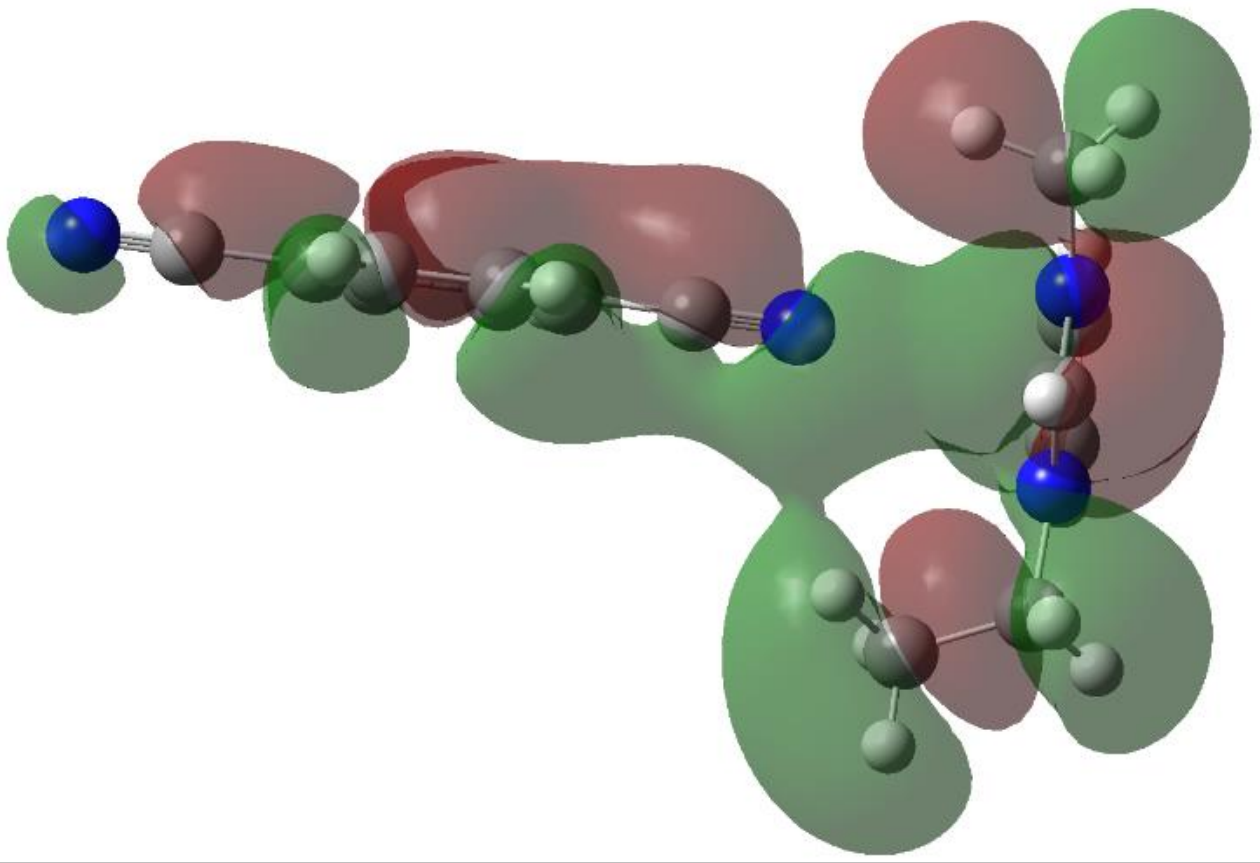 |
| **HOMO – 26**<br><br>Lone pair-π interactions of the nitrogen lone pair with the π electrons of the $[C_2C_1Im]^+$ ring. | 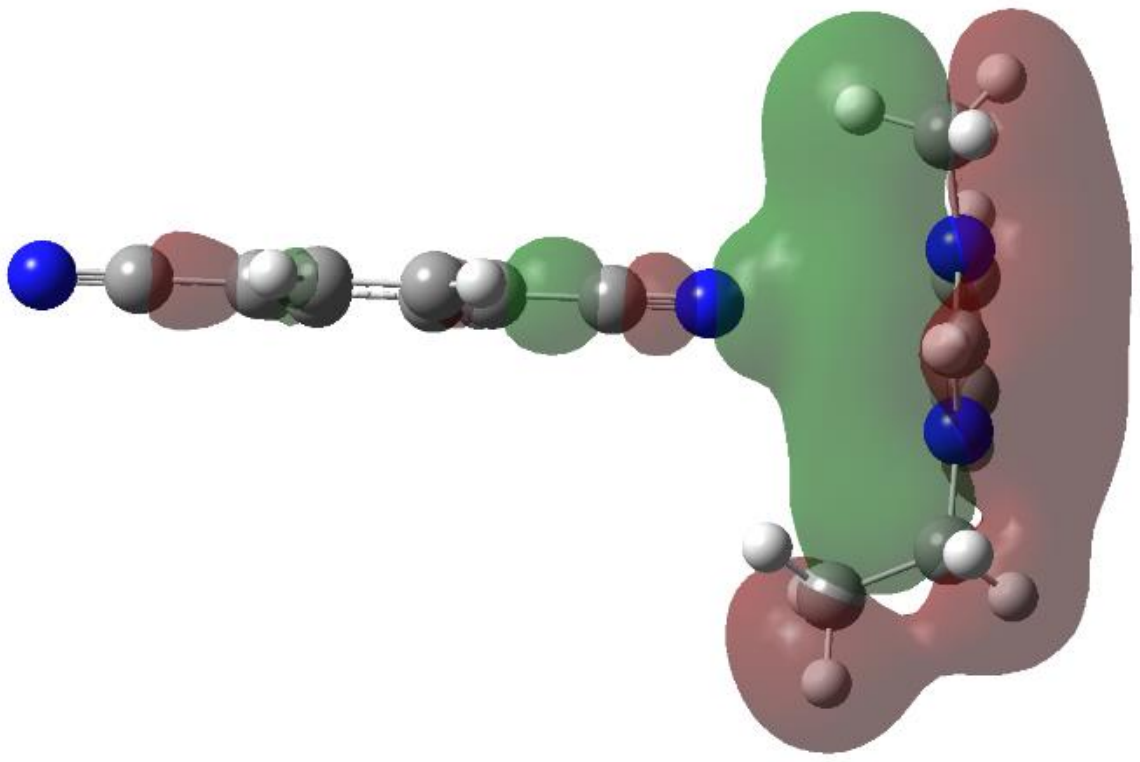 |
| **BzCN-[$C_2C_1Im$]$^+$ (4)**<br><br>Isovalue = 0.01 | 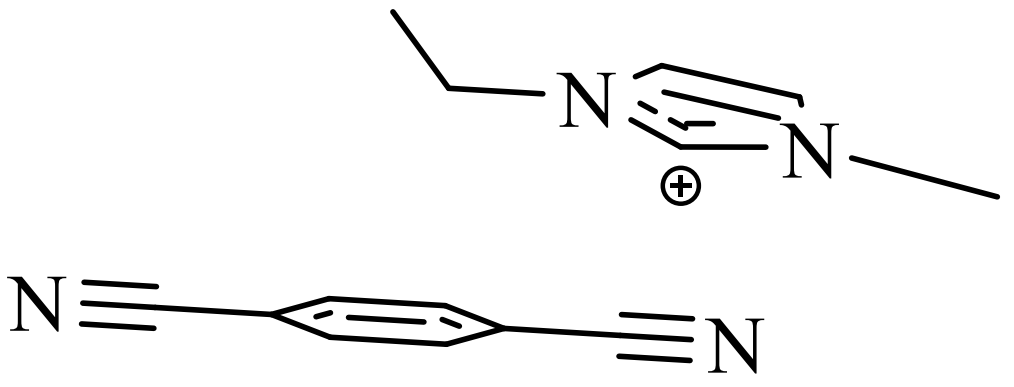 |
| **HOMO – 13**<br><br>Bonding interactions of nitrogen lone pair electrons or CN π electrons on BzCN and the π electrons in the ring of [$C_2C_1Im$]$^+$.<br><br>(Potential mixing of orbitals of the CN group involved in bonding, see FOs below.) | 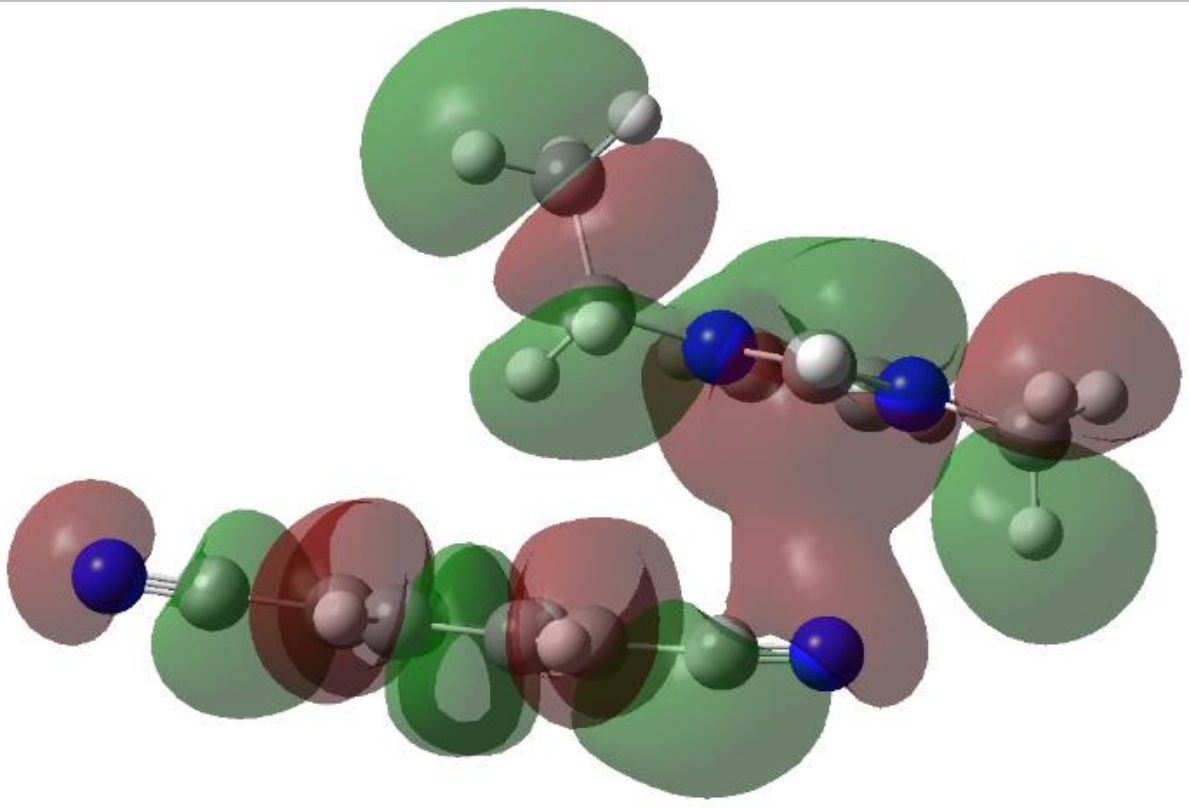 |
| **HOMO – 14**<br><br>Weak H-bonding between CN and hydrogen atom of the methyl in [$C_2C_1Im$]$^+$. | 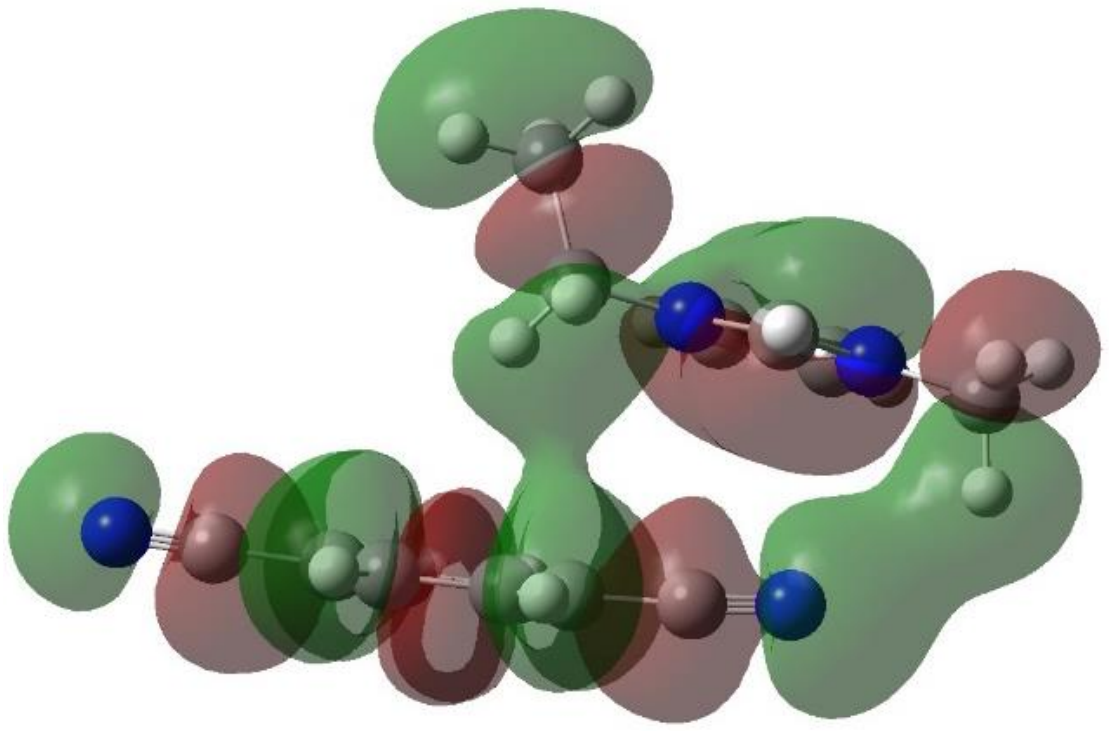 |
| **[$C_2C_1Im$]$^+$ FO for HOMO – 13**<br>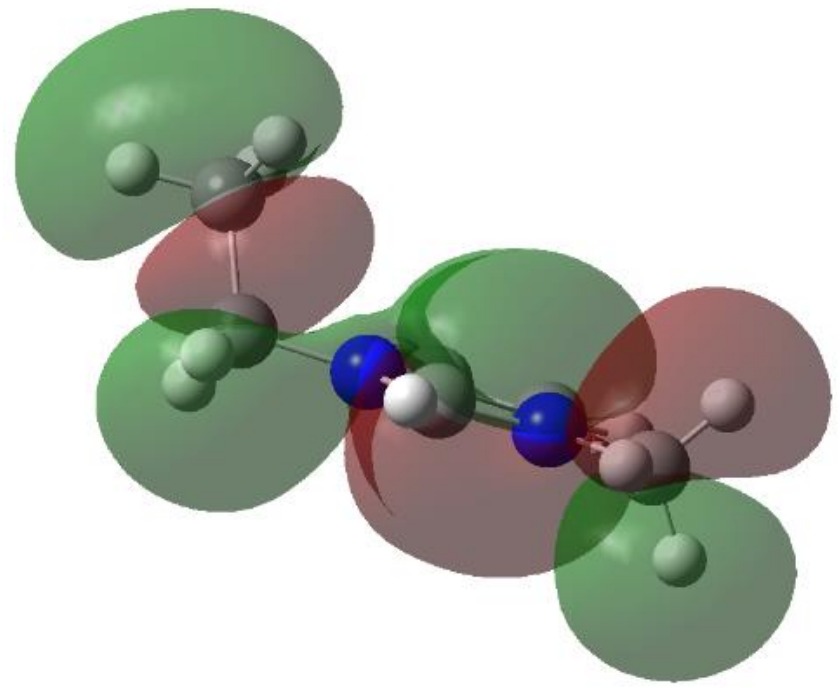 | **BzCN FO for HOMO – 13**<br>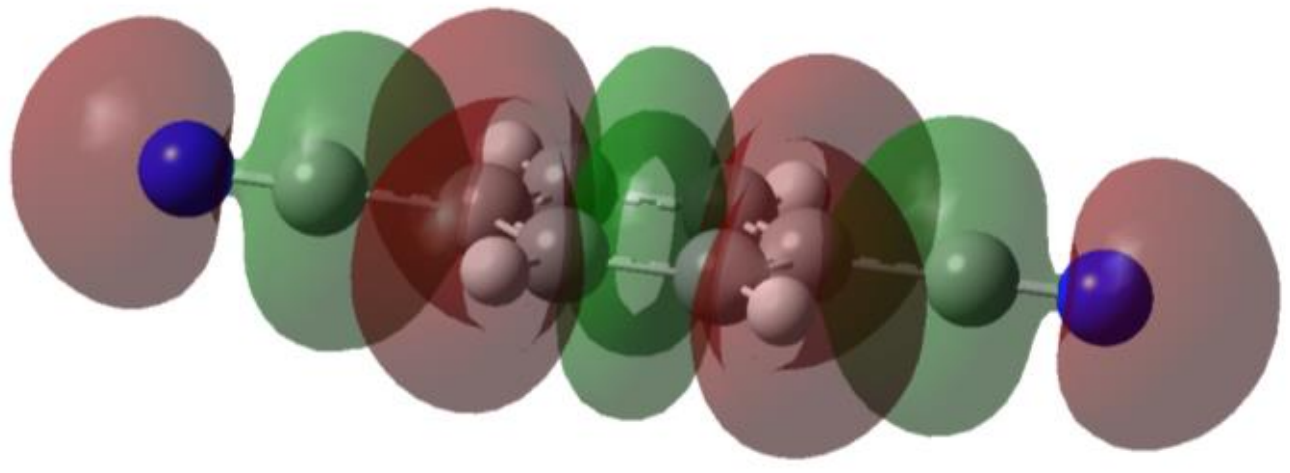 |

**Appendix.7: The π-π stacking distance in the most stable BzCC & $[C_2C_1Im]^+$ geometry.**

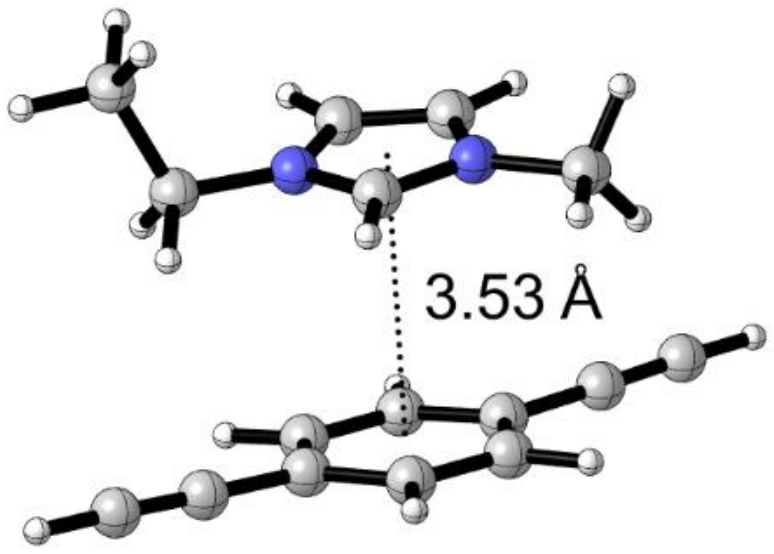

**Appendix.8: Selected bond length and angles for BzCC, BzCN, $[C_2C_1im]^+$ and $[MeSO4]^-$ in standard SMD and SMD-GIL environment.**

*Accuracy up to whole numbers for bond angles and 1 d.p. for bond lengths.*

| GP Analytes | Standard SMD | SMD GIL |
|---|---|---|
| BzCC | 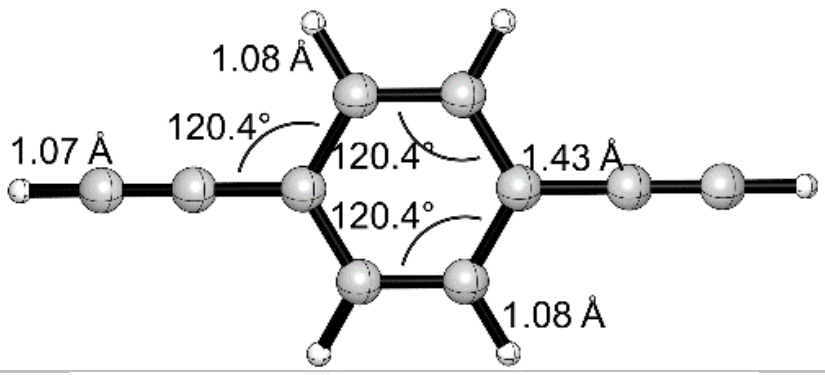 | 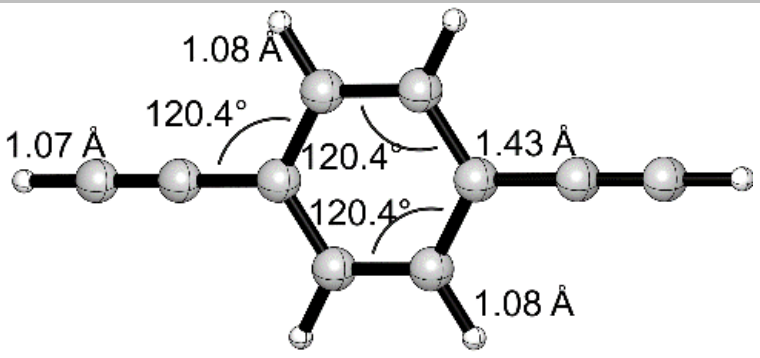 |
| BzCN | 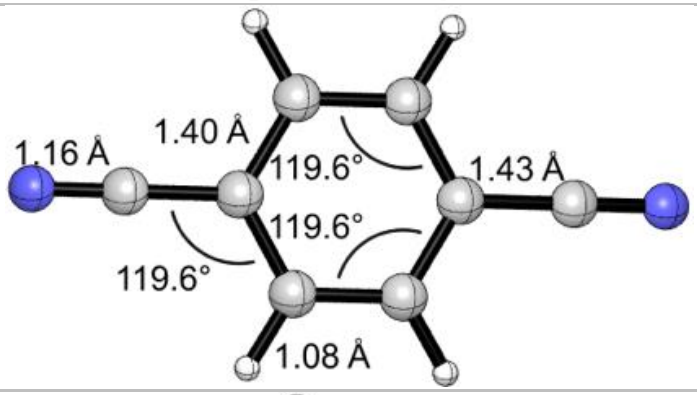 | 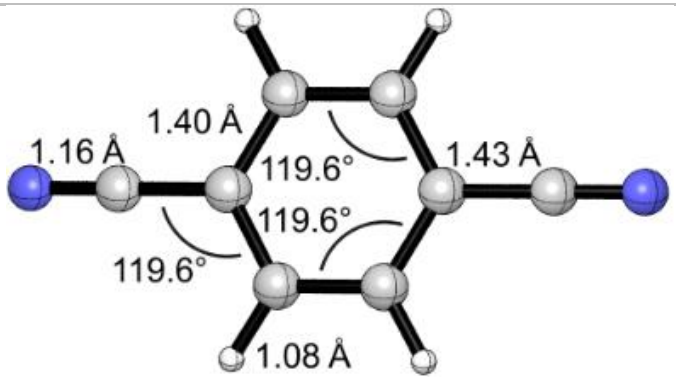 |
| $[C_2C_1im]^+$ | 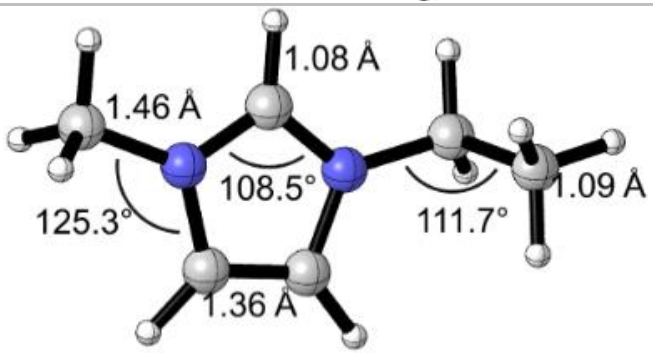 | 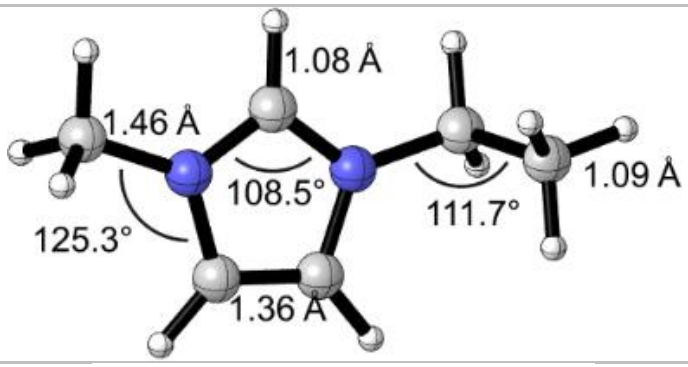 |
| $[MeSO4]^-$ | 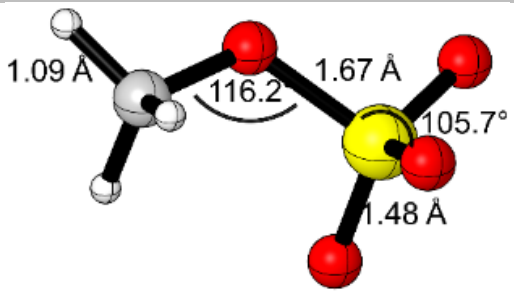 | 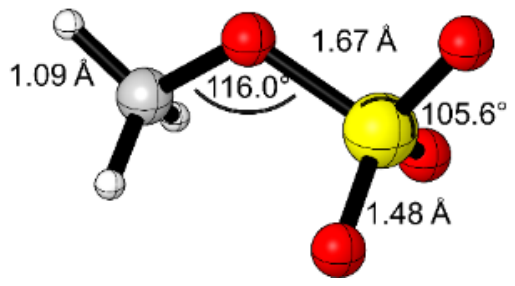 |